\documentclass{article}
\usepackage[utf8]{inputenc}
\usepackage{graphicx}
\graphicspath{{./image/}}
\usepackage{caption}
\usepackage{subcaption}
\usepackage{authblk}
\usepackage{lipsum}
\usepackage[dvipsnames]{xcolor}
\usepackage{bibentry}
\usepackage{bbm}
\usepackage{natbib}
\usepackage{tikz}
\usepackage{algorithm}
\usepackage{algpseudocode}
\usepackage[inline]{trackchanges} 
\usepackage{amsthm,amssymb,amsmath}
\usepackage{setspace}

\addeditor{FB}
\addeditor{FO}
\addeditor{ND}
\addeditor{TR}

\newtheoremstyle{theoremsstyle}
  {15pt}   
  {15pt}   
  {\itshape\setstretch{1.15}}  
  {}       
  {\bfseries} 
  {.}         
  {5pt plus 1pt minus 1pt} 
  {}       
\theoremstyle{theoremsstyle}
\newtheorem{theorem}{Theorem}[subsection]

\author[1]{Ferdinand Bhavsar}
\author[1]{Nicolas Desassis}
\author[1]{Fabien Ors}
\author[1]{Thomas Romary}
\affil[1]{Mines Paris, PSL University, Centre for geosciences and
geoengineering, 77300 Fontainebleau, France}
\usepackage[a4paper,
            bindingoffset=0.2in,
            left=1in,
            right=1in,
            top=1in,
            bottom=1in,
            footskip=.25in]{geometry}

\title{A stable deep adversarial learning approach for geological facies generation}
\date{}
            
\begin{document}

\maketitle 

\begin{abstract}
The simulation of geological facies in an unobservable volume is essential in various geoscience applications. Given the complexity of the problem, deep generative learning is a promising approach to overcome the limitations of traditional geostatistical simulation models, in particular their lack of physical realism. This research aims to investigate the application of generative adversarial networks and deep variational inference for conditionally simulating meandering channelized reservoir in underground volumes. In this paper, we review the generative deep learning approaches, in particular the adversarial ones and the stabilization techniques that aim to facilitate their training. We also study the problem of conditioning deep learning models to observations through a variational Bayes approach, comparing a conditional neural network model to a Gaussian mixture model.
The proposed approach is tested on 2D and 3D simulations generated by the stochastic process-based model Flumy. Morphological metrics are utilized to compare our proposed method with earlier iterations of generative adversarial networks. The results indicate that by utilizing recent stabilization techniques, generative adversarial networks can efficiently sample from complex target data distributions.
\end{abstract}

\section{Introduction}
\label{intro}

The spatial distribution of lithofacies in the subsurface is needed in a wide range of geoscientific applications such as petroleum engineering \citep{caers2005petroleum}, hydrogeology \citep{kitanidis1997introduction}, geophysics \citep{LINDE201586}, in situ recovery of minerals \citep{langanay2021uncertainty}, geothermal ressources assessment \citep{FOCACCIA201693}, carbon dioxide storage \citep{ZHONG201961}... where it is required as an input for the fluid flow simulations of physico-chemical processes. As the information about these properties is generally scarce and uncertain, coming from a few well logs or core drillings ("hard" data), undirect geophysical data or geological qualitative information about the area ("soft" data), geoscientists often resort to probabilistic models to account for possible scenarios and characterize the related uncertainties. These models need to meet several conditions to be useful in practice: they need to be fast to generate a great number of simulations of large 3D grids, honour the available data and exhibit geological consistency. They also necessitate being easily embedded into a data assimilation framework to incorporate other sources of data, e.g. fluid flow simulation results.

Geostatistics has aimed for decades to develop such models \citep{chiles}. Object based or boolean models \citep{chiu2013stochastic}, initially proposed to describe the distribution of grains and pores on a microscopic scale in reservoir rocks \citep{matheron1967elements} then materials \citep{jeulin2021morphological}, have been used at a macroscopic scale to characterize the main geological bodies such as channels and lenses in a reservoir \citep{dubrule1989review}. While powerful algorithms have been developed to generate conditional realizations of these models \citep{lantu},  the lack of flexibility induced by the parameterization of the objects makes them difficult to handle and their geological realism questionable. Later on, plurigaussian models have been proposed \citep{matheron1987conditional,aglge} to allow for more flexibility in the simulated bodies. The conditioning is made easy through the Gibbs sampler \citep{geman1984stochastic} but the inference of the model is not straightforward. In particular, the separation of the variability offered by the model between the facies proportions and the underlying Gaussian random function is not unique and hence must be cautiously carried out \citep{aglge}. Moreover, since the structure of the variability is driven by that of the latter two components, the realizations may fail to reproduce complex geological patterns. At the beginning of the century, the multiple-point approach \citep{streb} arose as a potential method to generate geologically realistic conditional realizations. Based on the reproduction of features found in a training image, this method allows to reproduce virtually any texture and has been successfully applied to a large variety of problems \citep{mariethoz2014multiple}. 

Meanwhile, sedimentologists have developed numerical models that attempt to reproduce the physics of the sediment deposition processes \citep{jacod}. As an example of such models, the Flumy Model \citep{flumy} generates 3D sedimentary deposits, taking into account processes such as avulsion, aggradation and migration. Due to their genetic nature, these models simulate realistic reservoirs. However, the generation is computer intensive. Furthermore, there exist, at the moment, no satisfactory way to condition the realizations to hard data \citep{bubnova:tel-02173727, troncoso:tel-04077499}.

Generative adversarial networks are a class of deep learning algorithms \citep{goodfellow2014generative} that exhibit great results in learning to generate fake realizations of complex multivariable distributions hardly distinguishable from real realizations \citep{arjovsky2017principledtraining} and have been applied in generating realistic pictures such as face images \citep{brock_large_2019, zhu_unpaired_2020} and art images \citep{karras2019style}. These models can be, despite these good results, hard to train, as discussed in \citet{arjovsky2017principledtraining} and research to improve the GAN method is still ongoing. Current improvements have come in the form of changing the architecture to generate images at different resolutions \citep{DBLP:progressive_growing_of_gans, msggan} or to use a Wasserstein loss \citep{arjovsky2017wasserstein, arjovsky2017principledtraining}.

Given their seemingly limitless possibilities, geoscientists are using these models increasingly. One of the first successes in the domain was the reconstruction of porous media, using a deep convolutional variant of the algorithm trained on segmented volumetric images of porous media \citep{mosser_porousmedia_gan}. Reproduction of semi-straight and meandering channelized structures was done with a convolutional Wasserstein GAN model \citep{chan_shing_gan_channelized}, and later research showed the models could do it while having the flexibility to generate conditional simulations \citep{dupont2018generating}. This method was extended to 3D volumes with a modified contextual loss \citep{zhang_application_2021}. Another method of conditioning applied Bayesian conditions, making use of Markov chain Monte Carlo (MCMC) methods to sample the posterior distribution. This method was successfully used with Variational Auto-Encoders models \citep{laloy_vave:hal-02094960} and later GANs \citep{laloy_gan}. Additionally, the progressive growing of GAN method also demonstrated good performance when used to generate geological facies \citep{progan_geol}.
Our objectives are to propose a deep generative model that fulfills the aforementioned requirements, addressing the shortcomings of geostatistical methods and numerical process-based models, while providing theoretical background and guidelines.

We first introduce deep adversarial networks, before expanding on their improved versions and the architecture details for non-conditional generation.
We then present two variational approaches to solve the conditioning problem, one from earlier literature using a neural network, and a new method using a Gaussian Mixture. Both of which can be used in 2D and 3D.
In the last section, we apply this model to generate Flumy-like stationary realizations of meandering channelized reservoir in two and 3D, conditioning them to pixel values, as well as showing the applicability of Conditional GANs (CGANs) to parameterized simulations, before concluding.

\section{Stable stationnary GAN}\label{descript_gan_archi}
In the past few years, Generative Adversarial networks \citep{goodfellow2014generative}, a sub-class of deep generative learning models, have been popular for being able to learn the approximation of complex, high-dimensional probability distribution using a large set of realizations thereof. These methods have seen a great amount of success in the generation of images such as pictures of faces and art \citep{arjovsky2017principledtraining, brock_large_2019, zhu_unpaired_2020, karras2019style}. Deep generative models have also increasingly been used to analyze spatial or spatio-temporal phenomena, making use of the flexibility of neural networks to learn complex data distributions \citep{mosser_porousmedia_gan, chan_shing_gan_channelized, laloy_gan, article_flumy_gan_chao}.

Generative Adversarial networks (GAN) consist of two separate neural networks, called the discriminator and the generator. These networks are trained by competing against each other. The goal is for the generator to sample synthetic data that are indistinguishable from training data, while the discriminator tries to distinguish between real and synthetic data.
The generator produces synthetic data $x$ by sampling from a Gaussian prior distribution and transforming it through a function $G(z,\theta^G)$, where $z$ is a realisation of the input distribution and $\theta^G$ are the parameters of the generator. Through $G$, the model thus defines the synthetic data distribution $p_\theta$. The generator tries to fool the discriminator into asserting its generated examples as real.
The discriminator assigns a probability value to each datum it receives as an input. Formally, it defines a family of functions $D(x, \theta^D)$ from the realisation space to the interval $[0,1]$, where $x$ is a data point and $\theta^D$ are the parameters of the discriminator. The closer $D(x, \theta^D)$ is to $1$, the higher the probability that $x$ was sampled from the training data distribution $p_r(x)$. A sigmoid function usually serves as its last activation function to ensure that the output of the discriminator is a valid probability.

The models are trained as a zero-sum game, in which the payoff of the discriminator is denoted $v(\theta^G, \theta^D)$ and that of the generator is $-v(\theta^G,\theta^D)$. Both models will attempt to maximize their own payoff, meaning that the optimal model for both will optimize the following min-max equation:
\begin{equation}
    \underset{\theta^G}{\arg\min}\ \underset{\theta^D}{\arg\max}\ v(\theta^G, \theta^D) = \underset{\theta^G}{\arg\min}\ \underset{\theta^D}{\arg\max}\ \mathbb{E}_{x\sim p_r} \log\ D(x) + \mathbb{E}_{x\sim p_{\theta}} \log(1-D(x))
\end{equation}

Despite the initial successes, early literature authors identified how strenuous it is to train GANs using the original architecture \citep{goodfellow2014generative}, with the need for extensive hyper-parameters tuning \citep{arjovsky2017wasserstein, arjovsky2017principledtraining}. The models may suffer from the vanishing gradient phenomenon \citep{article_vanishing_gradient}. Another suspected cause is the demonstrated fact that support of $p_r$ lies on low-dimensional manifolds \citep{manifold_hypothesis}, that $p_\theta$ is unlikely to cover at the early steps of the training. This makes the loss function saturated, which also causes the gradient to vanish.

A proposed strategy to make the support overlap more likely concentrate on reducing the effect of the curse of dimensionality. \citet{DBLP:progressive_growing_of_gans} tried to stabilise the outputs of the generator and speed-up convergence by making a GAN model progressively grow. They built more and more detailed image generators, starting with a $4 \times 4$ generator, and then adding layers during training. Intuitively, this helps with the overlap problem, since the distributions will  match first on lower resolutions where the overlap is more likely to be non-negligible.

Multi-Scale GANs \citep{msggan} relies on similar theoretical ideas to the progressive growing of GAN \citep{DBLP:progressive_growing_of_gans}, but diverge from the original method as the models do not grow throughout training. Instead, the models stay the same and learn at different scales from start to end. Thus the generator generates the realisation at different scales while the discriminator assesses the realism of the realisation at each scale. The real and generated realisations at different scales are concatenated through these connections to the activation features of the discriminator at the corresponding scale.

The advantage over the progressive growing of GANs is a simpler implementation, as well as allowing gradient circulation in the model at all scales all along the training iterations, preventing the vanishing gradient problem.



Instead of making the overlap more likely, another popular method avoids the problem completely by resorting to a different loss function. The Wasserstein-1 distance, a distance measure between two probability distributions, has been proposed in \citet{arjovsky2017wasserstein}. This loss is computed using its dual form named Kantorovich-Rubinstein dual (see Annex \ref{appendix_kr_dual} for details). This approach requires however the discriminator to be 1-Lipschitz (Annex \ref{appendix_kr_dual}). To enforce this condition, we use a Spectral Normalization \citep{spectral_normalisation}:
\begin{equation}
    W_{SN} = \frac{W}{\sigma(W)}
\end{equation}
where $\sigma(W)$ is the spectral norm of parameters $W$.

 We also change the activation functions following the recommendations made by \cite{sorting_lip_cemanil}. We use non-monotonic, 1-Lipschitz bounded functions or gradient preserving functions such as GroupSort, Fpflu  \citep{FPFLU} and the Swish activation functions \citep{ramachandran2018searching} normalized to have unit gradient norm. Furthermore, to ensure that our model is entirely translation invariant, as we need to model stationary processes, our models are entirely convolutional. This aspect is similar to previous work on stationary GANs \citep{spatial_gan, long2015fully, laloy_gan}. The detailed architecture is described in Section \ref{section_application}.

\section{Conditioning the generator to observations}
\label{model_mswgan}
Our objectives are not only to propose a stable deep generative model that simulates realistic reservoirs, but also a way to condition this generator  to observations. We describe here two competing approaches based on a variational Bayes framework.
\subsection{Variational Bayes Conditioning}

Once trained to optimality, the GAN framework provides a generator $G$ that can simulate new realisations with the same properties as the training dataset, as such we can consider that using the generator samples $x$. Applications, in resource prospecting for example, require sampling realizations knowing partial information about $x$. In geology, this partial information can come from boreholes or seismic imaging. We denote the partial information $x^\star$. Its conditional distribution with respect to $x$, also called observation process, is modeled by specifying the conditional density $p(x^\star|x)$. When generating realizations through the generator, $x = G(z)$, with $z$ the input of the generator, we can instead consider $p(x^\star|z)$, where $x^\star$ is a set of $N$ observations. We recall that $z$ is a standard Gaussian random vector with independent components, that defines the prior distribution $p(z)$. Therefore, the conditioning problem consists in obtaining the distribution $p(z|x^\star)$ by using the Bayes formula:
\begin{equation}
p(z|x^\star) = \frac{p(x^\star|z)p(z)}{p(x^\star)}\\
\end{equation}
\begin{equation}
p(x^\star) = \int_{} p(x^\star|z)p(z) dz
\end{equation}

The computation cost of this integral is exponential in the dimension of the problem. Consequently, instead of computing it exactly, we will use methods that allow us to ignore the normalizing constant. Monte-Carlo Markov Chains (MCMC) methods are usually used to sample from such a distribution. Previous methods have successfully applied MCMC complementary to GANs for geological simulations to sample conditional distributions \citep{laloy_gan}. However, they can be slow to converge and may encounter difficulties to explore several posterior modes.

Variational methods change the sampling problem into an optimisation problem, which is solved through gradient descent. To do so, we limit our research to a family of parametric distributions $\mathcal{Q}$, defined on the same probability space as $p(z|x^\star)$, the real latent  distribution. We search for the member of this family that minimises the \textbf{Kullback-Leibler Divergence} between itself and the real distribution:
\begin{equation}
q_\theta^{*}(z|x^\star) = \operatorname*{\arg\min}_{q_\theta(z|x^\star) \in \mathcal{Q}} D_{KL}\Big(q_\theta(z|x^\star)\ ||\ p(z|x^\star)\Big)
\end{equation}

This new form of the conditioning problem allows the use of common optimisation methods, to optimise the parameters $\theta$ of the parametric distribution $q_\theta$. Moreover, once trained, the variational approximation allows to generate any number of conditional realisations.

\subsection{Inference Network}\label{sec:conditioningalgo}
In \cite{chan_shing_gan_channelized}, the authors use an inference neural network $I$, such that the family of parametric distribution $\mathcal{Q}$ is described by the set of all the functions defined by the neural network $I$ applied to a given set of random variables. According to the Universal Approximation Theorem \citep{pinkus_1999, DBLP:journals/corr/abs-1905-08539}, by choosing a neural network deep enough, we can learn an approximation of the true posterior distribution $p(z|x^\star)$.
\begin{center}
\includegraphics[width=\linewidth]{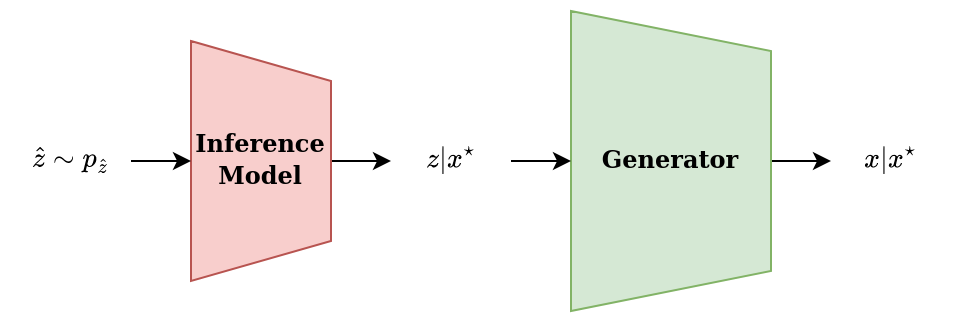}
\captionof{figure}{Following the method from \cite{chan_shing_gan_channelized}, we use a secondary network that will transform a random vector $\hat{z}$ into our conditional posterior $z|x^\star$.}
\end{center}

The neural network defines the parametric distribution $q_\theta$ by transforming $\hat{z}$, a standardized Gaussian vector with independent components, into $z|x^\star$. The model is a simple Multi-Layer Perceptron, with dense layers and Selu activation functions \citep{DBLP:journals/corr/KlambauerUMH17}. The input latent space and output space are of the same dimension.

The inference neural network $I$ is parameterized by $\theta$, which are the weights of the linear layers. It needs to be trained such that its weights $\theta$ are optimal. The optimal weights $\theta^*$ minimise the Kullback-Leibler divergence between the parametric distribution and the real posterior distribution:
\begin{equation}
\begin{aligned}
\theta^* = &\operatorname*{argmin}_{\theta} D_{KL}\Big(q_\theta(z|x^\star)\ ||\ p(z|x^\star)\Big)\\
& = \operatorname*{argmin}_{\theta}\ \mathbf{H}(q_{\theta},p) - \mathbf{H}(q_{\theta}), 
\end{aligned}
\label{eq:dkl_qp}
\end{equation}
where $\mathbf{H}(q_{\theta},p)$ is the negative cross-entropy between the training data distribution $p$ and the variational distribution $q_\theta$. In our application, the observations $x^\star$ are vertical wells, which are locations in the realisation where the category of facies is known. If we assume that observations are independent realizations of multinomial variables with probabilities given by the generator, we can compute the negative cross-entropy by using the multi-class cross-entropy. $\mathbf{H}(q_\theta)$ is the Shannon Entropy of the probability distribution $q_\theta$.
Hence, minimizing equation \eqref{eq:dkl_qp} amounts to minimizing the cross-entropy between the two distributions while keeping the entropy of the variational distribution high. The latter ensures that the model maximises the variability in the posterior distribution.

In our application, the observations are categorical vector of facies indicators.
Therefore, we make the hypothesis of independent multinomial observations. Because of this hypothesis and because $p(z)$ is the multivariate standard normal distribution, we can use Bayes' rule to compute $\mathbf{H}(q_{\theta},p)$ as follows: 
\begin{equation}
\begin{aligned}
    \mathbf{H}(q_{\theta},p)  & = \mathbb{E}_{z\sim q_\theta} -\log p(z|x^{\star})\\
    & =  \mathbb{E}_{z\sim q_\theta} \Big[-\sum_{i=1}^{N} \log p(x_i^{\star}|z) + \frac{1}{2}||z||^2\Big]
\end{aligned}
\end{equation}
with $N$ the number of individual conditioning observation points $x_{1\leq i\leq N}^{\star}$. The expectation is intractable, because we can only sample $p(x_i^{\star}|z)$ through the generator, as such we use a Monte-Carlo estimator over a batch of $M$ generated $z_m = I(\hat{z})$:
\begin{equation}
\begin{aligned}
    \mathbf{H}(q_{\theta},p)  & =  \frac{1}{M} \sum_{m=1}^{M} \Big[-\sum_{i=1}^{N} \log p(x_i^{\star}|z_m) + \frac{1}{2}||z_m||^2\Big]
\end{aligned}
\end{equation}
The second term of \eqref{eq:dkl_qp}, namely the entropy $\mathbf{H}(q_{\theta})$ requires also an approximation since we have access to $q_{\theta}$ only through its realisations. $\mathbf{H}(q_{\theta})$ can be approximated through the Kozachenko-Leonenko estimator \citep{kozachenkoleonenko}, see Appendix \ref{appendix_kozachenko_leonenko} for more details. The approximate loss function is hence defined as follows:
\begin{equation}
D_{KL}\Big(q_\theta(z|x^\star)\ ||\ p(z|x^\star)\Big)
\approx -\frac{d}{M} \sum_{m=1}^{M} \log(d_{knn}(z_m)) + \frac{1}{M} \sum_{m=1}^{M} \Big[-\sum_{i=1}^{N} \log p(x_i^{\star}|z_m) + \frac{1}{2}||z_m||^2\Big]
\end{equation}
with $d_{knn}(z_m)$ the distance from sample $z_m$ to its $k^{th}$ neighbour. Following \cite{chan_shing_gan_channelized}, we set $k = \sqrt{M}$.

\subsection{Mixture of Gaussians}\label{gaussian_mixture_part}
The previous approach may suffer from mode collapse, which is an optimisation error where the algorithm only approximates one mode of a multimodal distribution. Indeed, the Kullback-Leibler divergence is blind to isolated modes \citep{blindnessmodes}. We propose to alleviate this problem by using a mixture of Gaussians in the variational approach.

We choose a number of Gaussian components $K$. Each component $i=1,..., K$ is defined by its mean $\mu_i$ and its covariance matrix $\Sigma_i$, as well as non-negative weights $\pi_i$ that sum to $1$. The parametric distribution is then defined as $q_\theta(z) =\sum_{i=1}^K\pi_i f_{\mu_i,\Sigma_i}(z)$, where $f_{\mu,\Sigma}$ is the gaussian pdf with mean $\mu$ and covariance $\Sigma$.

Similarly to the inference network method, a standardized Gaussian vector with independent components $\hat{z}$ is sampled. Then a random component of the mixture $i$ is chosen using the mixture weights. The mean and covariance of the chosen component are finally applied to the initial standard Gaussian random vector with independent components:
\begin{equation}\label{mixture_gaussian_equation}
    z = \mu_i + \Sigma_i^{1/2} \times \hat{z}, 
\end{equation}
where $\Sigma_i^{1/2}$ is e.g. the Cholesky decomposition of $\Sigma_i$.

To train the parameters of the components we again use the Kullback-Leibler divergence as the objective. This means that we cannot guarantee that it will find every modes of the distribution. However, in our experiments, we have found that by selecting enough components in the mixture and initializing them randomly, this method exhibits a greater degree of diversity compared to the deep inference model. Choosing a large number of components increases the complexity of the model, as well as the time it takes to train it.

The Kullback-Leibler divergence is decomposed in the same manner as above. We keep $\theta$ as the notation for parameters, noting that $\theta_i = (\pi_i, \Sigma_i, \mu_i)$ and $\theta = (\theta_1, \theta_2, ... \theta_K$). Since we are working with Gaussian Mixtures we can compute the entropy using the approximation proposed by \cite{entropy_mixture_gaussian}:
\begin{equation}
    \begin{aligned}
        \mathbf{H}(q_\theta)  & \approx \mathbf{\Tilde{H}}(q_\theta)\\
        & = -\sum_{i=1}^K \pi_i \int \mathcal{N}(z|\mu_i, \Sigma_i)\log(\pi_i\mathcal{N}(z|\mu_i,\Sigma_i))dz\\
        & = \frac{m}{2} + \frac{m}{2} \log2\pi +\frac{1}{2}\sum_{i=1}^K\pi_i\log|\Sigma_i|-\sum_{i=1}^K\pi_i\log\pi_i
    \end{aligned}
\end{equation}
The cross-entropy is also weighted by the components used to generate the simulations of the current batch. During an iteration, we sample $M$ generated noises $z_m$. Each $z_m$ is generated by sampling one random mixture component with parameters $(\pi_m, \Sigma_m, \mu_m)$. Finally, the Kullback-Leibler divergence is approximated as follows:
\begin{equation}
D_{KL}\Big(q_\theta(z|x^\star)\ ||\ p(z|x^\star)\Big)
 \approx  -\mathbf{\Tilde{H}}(q_\theta) + \frac{1}{M} \sum_{m=1}^{M} \pi_m \Big[-\sum_{i=1}^{N} \log p(x_i^{\star}|z_m) + \frac{1}{2}||z_m||^2\Big]
\end{equation}

\section{Application}\label{section_application}
In this section, we apply the generative models described earlier, to produce reservoir simulations generated with Flumy, a process-based model for meandering channelized reservoir, similarly to \cite{article_flumy_gan_chao}.
Every model and metric used for our experiments are coded in Python and the Tensorflow library is used to design the deep learning models.
The models are trained with Graphical Processing Units from Google Colaboratory resources.
In order to evaluate the quality of the simulations generated by the different GAN architectures tested, we compute and compare morphological properties such as facies proportions and connected components size distribution between GAN simulations and training dataset. 
The gstlearn \citep{gstlearn_software} Python package from MINES Paris is used for this purpose. We then test the best model on the task of conditioning.

\subsection{Flumy}
Flumy is a stochastic process-based model used to simulate the geometry of heterogeneous reservoirs generated by meandering channelized systems in both fluvial and turbidites environments \citep{lopez_flumy_2003, flumy}. Flumy combines first-order hydraulic equations to track the evolution of channels and stochastic algorithms simulating chute cutoff, levee breaching, avulsions and associated deposits. Flumy generates 3D reservoir models following a user-defined multi-sequence scenario.

\begin{center}
\includegraphics[width=\linewidth]{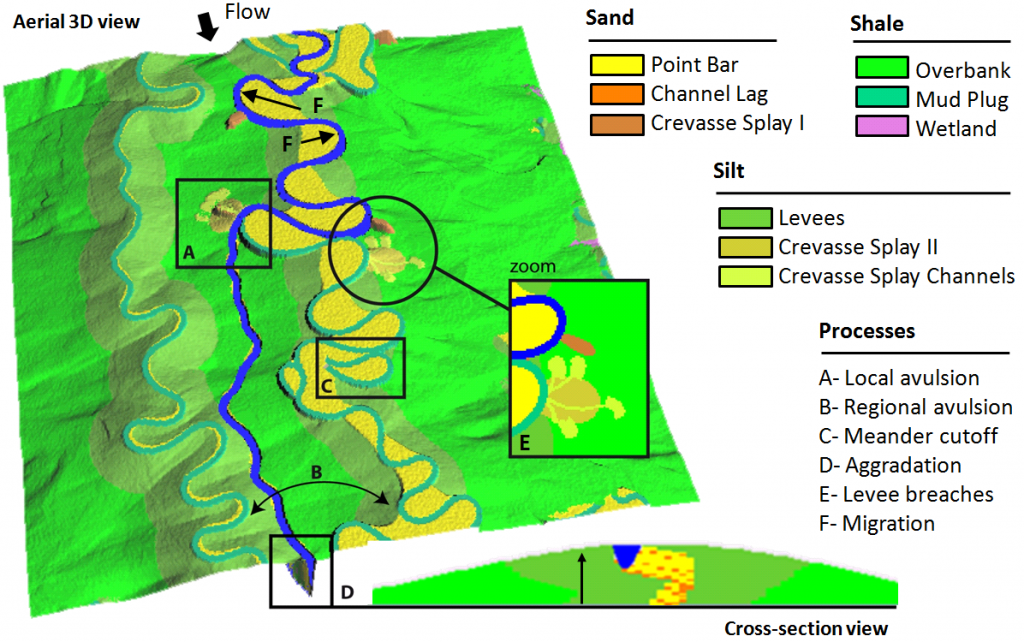}
\captionof{figure}{Overview of Flumy's main simulated processes for Fluvial systems \citep{flumy}}
\end{center}

Conditioning to existing subsurface datasets, while possible \citep{bubnova:tel-02173727, troncoso:tel-04077499}, is difficult and the method lacks of flexibility. Furthermore, inference computational cost is high, making it a choke-point for workflows where realisations must be simulated many times.

\subsection{Experimental setup}

Experiments are conducted using a dataset generated using the Flumy model. See Annex \ref{data_gen_process} for more details on the Flumy parameters used to generate the dataset.

The 2D GAN models are trained using a dataset containing 5000 horizontal slices extracted from a Flumy reference simulation with 4 facies (Channel Lag, Point Bar, Levee and Overbank) at a 10m resolution (2D realisation size is 64x128 pixels). The 3D GAN models are trained using a dataset containing 3000 3D blocks of similar meandering channelized reservoirs extracted from the Flumy reference simulation with a vertical discretization step of 10m with the same 4 facies at a 10m resolution (3D realisation size is 16x32x64 pixels). Realisations are represented as matrices, and at each position, the type of geological facies is a categorical vector of facies indicators. Consequently, in our dataset, there are only 4 different values a pixel can take (i.e. [1,0,0,0], [0,1,0,0], [0,0,1,0] or [0,0,0,1]).

Later in the application we test a CGAN model \citep{mirza2014conditional} for parameterized simulations. For this part of the application, we use other Flumy simulations which are variations around the default Fluvial scenario (see \cite{flumy}). Each Flumy simulation is tuned by setting a specific value of the two following key parameters: the required proportion of sand (aka. Net to Gross or NG) and the required mean Sand Body Extension Index (ISBX). Then, the 2D CGAN model is trained using a dataset of horizontal slices extracted from the different parameterized Flumy simulations with 9 facies (Channel Lag, Point Bar, Sand Plug, Crevasse Splay I, Crevasse Splay II Channels, Crevasse Splay II, Levee, Overbank, Mud Plug) at a 10m resolution (2D realisation size is 64x128 pixels).

\subsection{Meandering channelized reservoir simulations results}

\subsubsection{2D}
Initially, we test the Multi-Scale Wasserstein GAN with Spectral Normalization and GroupSort, whose architecture is given in appendix \ref{appendix:MSWGAN_arcitecture} on 2D simulations. In these tests, the generator architecture stays mostly the same, with two $(3,3)$ convolution operations at each level, LeakyReLU activation functions, and a final Softmax activation function. Furthermore, multi-scale generators require a pixelwise skip-connection at each level too. The critic architecture reflects the one of the generator, with two $(3, 3)$ convolutions operations at each levels. To reflect the skip connection in the generator, our critic returns a score at each level, with a $(3,3)$ convolution from which the mean value is our level score. The GroupSort operation is used everywhere after every convolution in the critic, with the exception of the outputs convolution.

For comparison purposes, we also train a Multi-Scale GAN with the vanilla GAN loss, and only the addition of a Gaussian noise to stabilize it. We also train two vanilla Wasserstein GAN with a Gradient Penalty, one with the GroupSort activation and the second without. The results presented are obtained with the set of hyper-parameters shown in the Table \ref{tab:hyperparameters2d}.

\begin{table}[h!]
    \centering
    \begin{tabular}{|c|c|c|c|c|}
        \hline
         2D Model & W-GAN-GP & MS-GAN & W-GAN-GP-gs & MW-GAN-SN-gs\\
         \hline\hline
         Epochs & 150 & 100 & 200 & 200\\
         lr gen./disc. & $5e^{-4}/1e^{-4}$ & $5e^{-4}/5e^{-4}$ & $5e^{-3}/5e^{-4}$ & $5e^{-3}/5e^{-4}$ \\
         \hline
    \end{tabular}
    \caption{Hyper-parameters used to train our showcased 2D models. Here "lr gen./disc." refers to the two learning rates used respectively for the generator and the discriminator (or critic).}
    \label{tab:hyperparameters2d}
\end{table}

Given the subjective nature of human visual assessment, we chose to quantify the quality of the non-conditional simulations using morphological metrics calculated from 1000 simulations per model. The distribution of connected component sizes by facies for multiple 2D models is depicted in Figure \ref{fig:connectedcomponentssize2d} and compared to the training data. A Wasserstein-1 loss is computed, between each of these distributions and the real distribution (Table \ref{tab:connected_components2d}). The mean proportion of each geological facies for the 2D models is shown in Figure \ref{fig:2d_proportions}. We can see on these figures that the distribution of smaller features ("Sand, Channel lag", "Silts, Levees" and "Shale, Overbank") are correctly predicted for most models except for the Wasserstein GAN with gradient penalty. However, the "Sand, Point bar" prediction has a very high loss value as seen in Table \ref{tab:connected_components2d}. This a phenomenon for which we did not find an explanation.

The metrics comparison for the different models shown in Figure \ref{fig:connectedcomponentssize2d} and Figure \ref{fig:2d_proportions} suggest that models utilizing Wasserstein or Multi-Scale techniques are able to generate simulations that are visually convincing. Morphological properties analysis, as shown in Figure \ref{fig:connectedcomponentssize2d}, also indicates that most of the models are able to capture the characteristic features of the input dataset. In this test, distribution of connected components size of each geological facies for different models are compared against training data. We can see from the comparison that our proposed method MSWGAN-SN is only the second best when reproducing the morphology of most facies except the "Sand, Point bar". We acknowledge that in 2D, a simple model seems more efficient. However, when we started to generate 3D blocks (see next section), we were not able to reach the convergence while training any other model except the one we introduced just before.

Furthermore, the model generates unique simulations which are not copies from the dataset. This can be seen in Figure \ref{figurevariation}, where we compare visually generated realisations to their nearest neighbour, in the Euclidean space, from the Flumy dataset.

\begin{center}
    \includegraphics[width=\linewidth]{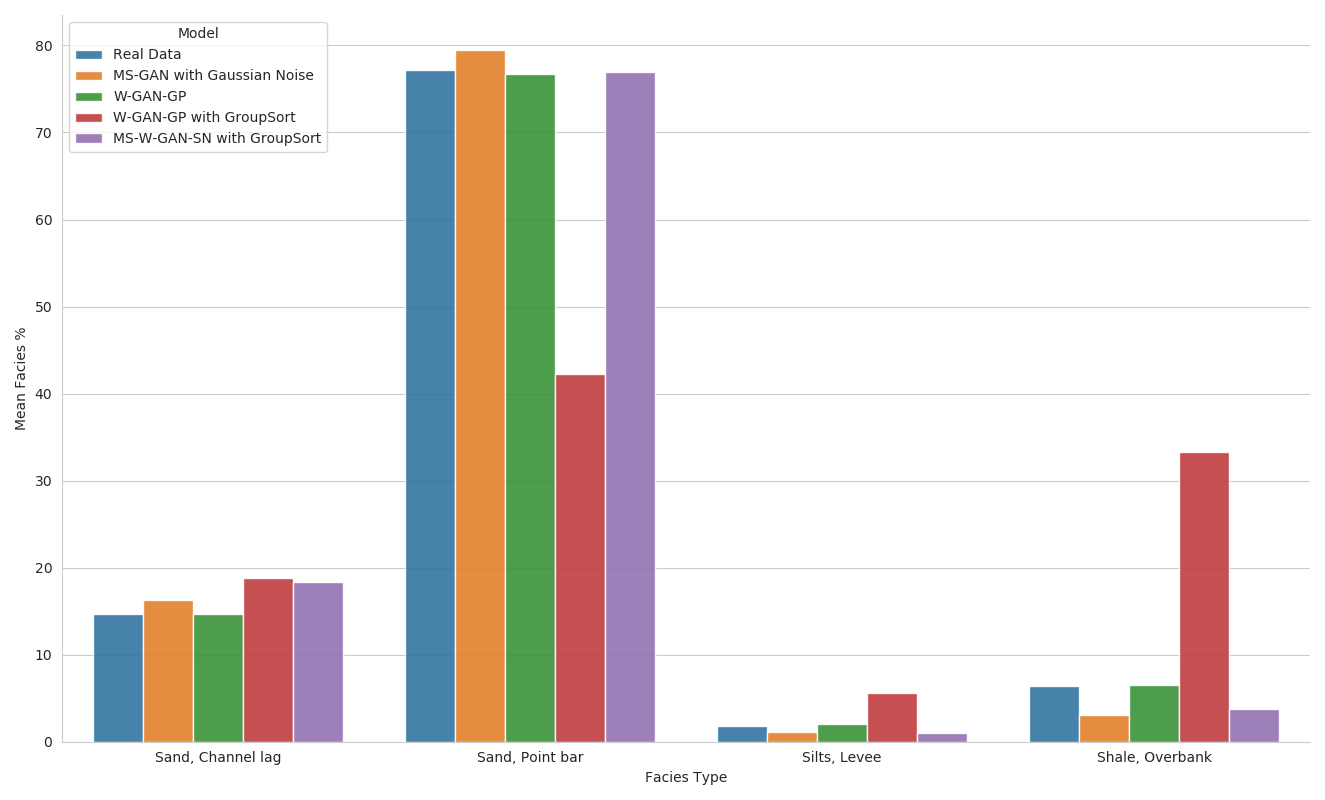}
    \captionof{figure}{Comparison of the proportions of different facies for our 2D models.}
    \label{fig:2d_proportions}
\end{center}

\begin{center}
    \includegraphics[width=\linewidth]{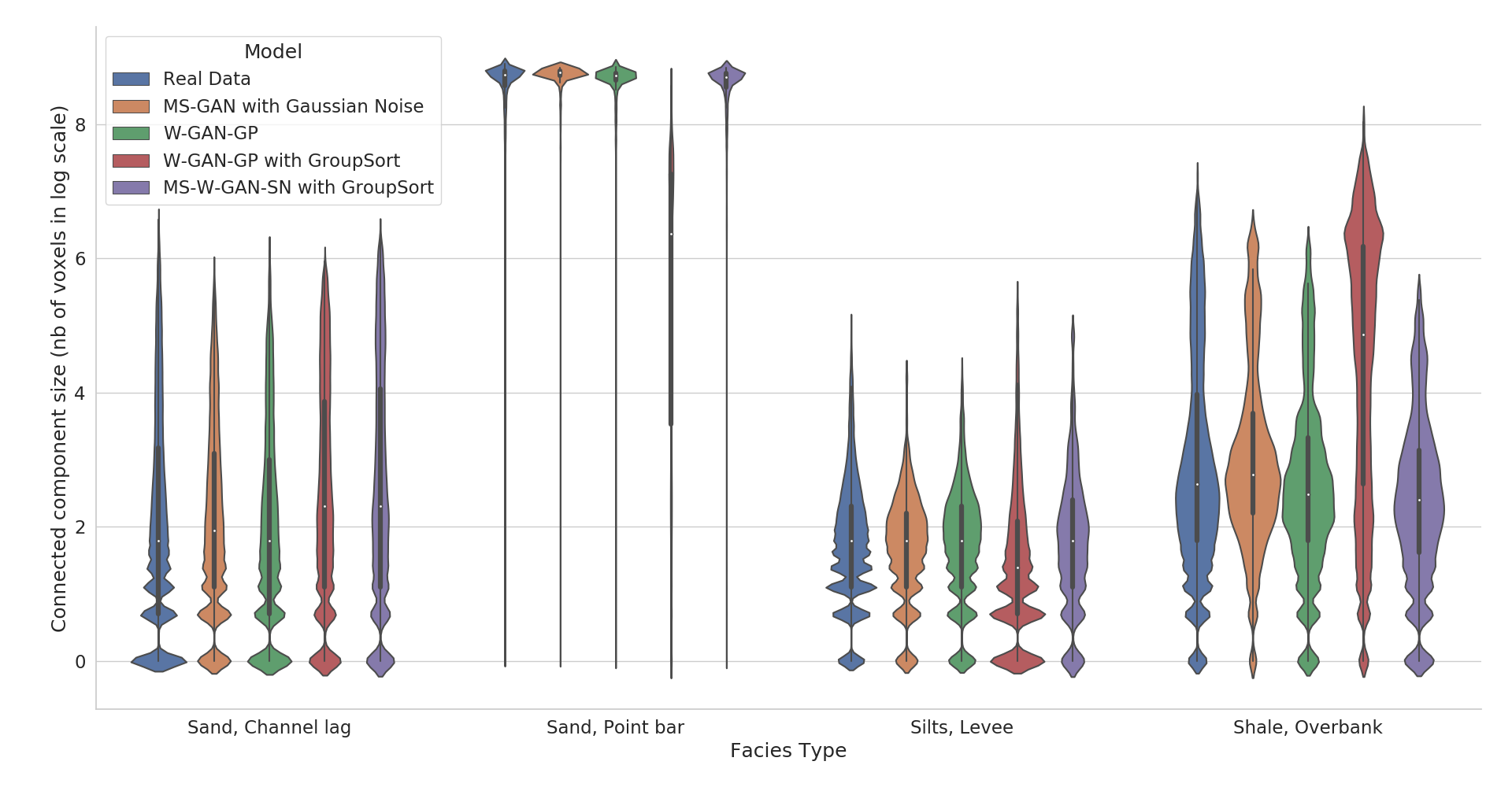}
    \captionof{figure}{Distribution of connected components size of each geological facies for different models against training data. To make visualisation more relevant, each connected component size is itself weighted by it's volume.}
    \label{fig:connectedcomponentssize2d}
\end{center}

\begin{center}
\includegraphics[width=0.5\linewidth]{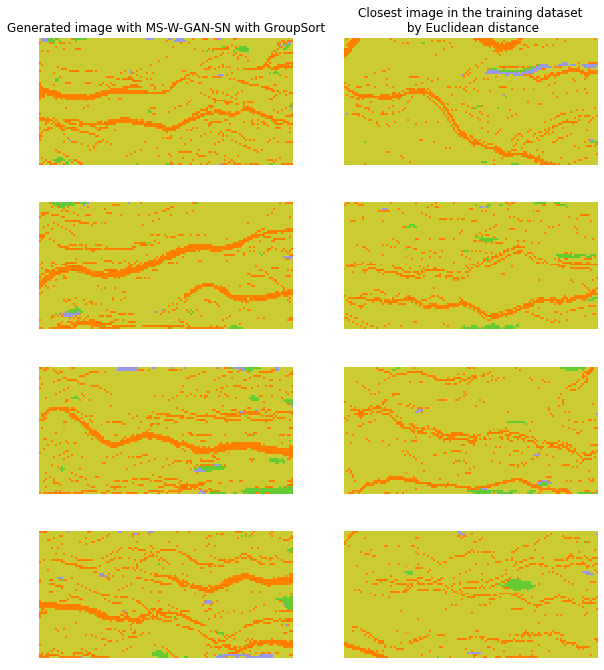}

\captionof{figure}{Comparison of the generated realisations (on the left) to their nearest counterparts in the training dataset (on the right) using Euclidean distance.}\label{figurevariation}
\end{center}

\begin{table}
\begin{center}
\begin{tabular}{| c || c | c  | c | c |}
 \hline
 Model & Sand, Channel lag & Sand, Point bar & Silts, Levee & Shale, Overbank \\ 
\hline
 MSGAN Gaussian Noise & 0.64 & 126.64 & 0.39 & 2.67 \\
\hline
 WGAN Gradient Penalty & 0.11 & 16.47 & 0.29 & 0.90 \\
\hline
 WGAN GP with GroupSort & 0.75 & 143.33 & 1.39 & 6.09 \\
\hline
 MSWGAN SN with GroupSort & 0.97 & 14.21 & 0.33 & 1.68 \\
\hline
\end{tabular}
\end{center}
\caption{To compare the connected components size distribution in 2D, shown in Figure \ref{fig:connectedcomponentssize2d}, a Wasserstein loss is computed between the model distributions and the training data distribution.}
\label{tab:connected_components2d}
\end{table}

\subsubsection{3D}
Once established the robustness of the 2D model, we extend the analysis to 3D simulations (Figure \ref{fig:3d_data_comparison}). The architecture design of the 3D model is identical to 2D architecture, with the exception that the 2D convolutions becomes 3D convolutions and the activation function is a normalized Swish. The hyper-parameters used to train 3D models are showcased in Table \ref{tab:hyperparameters3d}.
\begin{table}[h]
    \centering
    \begin{tabular}{|c|c|}
        \hline
         3D Model & MW-GAN-SN-gs\\
         \hline\hline
         Epochs & 200\\
         Learning rate generator/discriminator & $1e{-3}/1e{-3}$ \\
         \hline
    \end{tabular}
    \caption{Hyper-parameters used to train our showcased 3D model}
    \label{tab:hyperparameters3d}
\end{table}

We evaluate the 3D model using the same metrics as in 2D. Those are presented in Figure \ref{fig:3d_connected_components} and in Figure \ref{fig:3d_proportions}.\\

Training 3D models proved to be more challenging, and ultimately only a combination of Wasserstein and Multi-Scale techniques was successful. Both visual and morphological evaluations of the resulting 3D model demonstrate its ability to generate geological structures that are consistent with the input dataset. The visual evaluation, shown in Figure \ref{fig:3d_data_comparison}, reveals geological patterns found both in the Flumy simulations and the GAN simulations, specifically, the "V" pattern of sand in vertical sections and the elongated sand channels in horizontal slices. We also can notice the small scattered shale lenses in both Flumy simulations and GAN simulations. This is a good sign that the models captures the features from the input dataset. However, we can notice that the generated sand channels are not as horizontally continuous as the ones generated by Flumy.

\begin{center}
    \includegraphics[width=\linewidth]{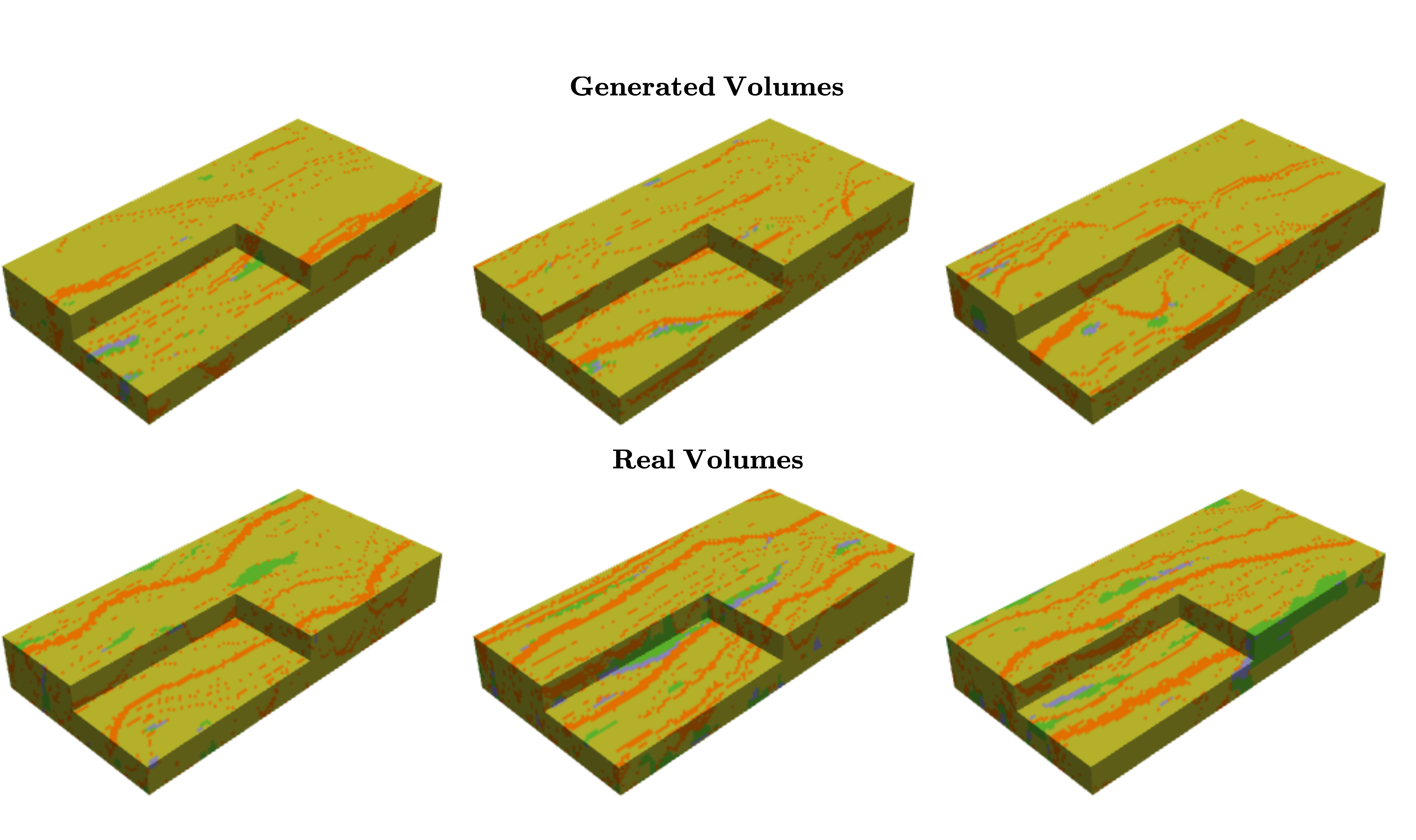}
    \captionof{figure}{Visual comparison of 3D volumes generated with our Multi-Scale Wasserstein GAN (top) and the Flumy 3D volume (bottom)}
    \label{fig:3d_data_comparison}
\end{center}

\begin{center}
    \includegraphics[width=\linewidth]{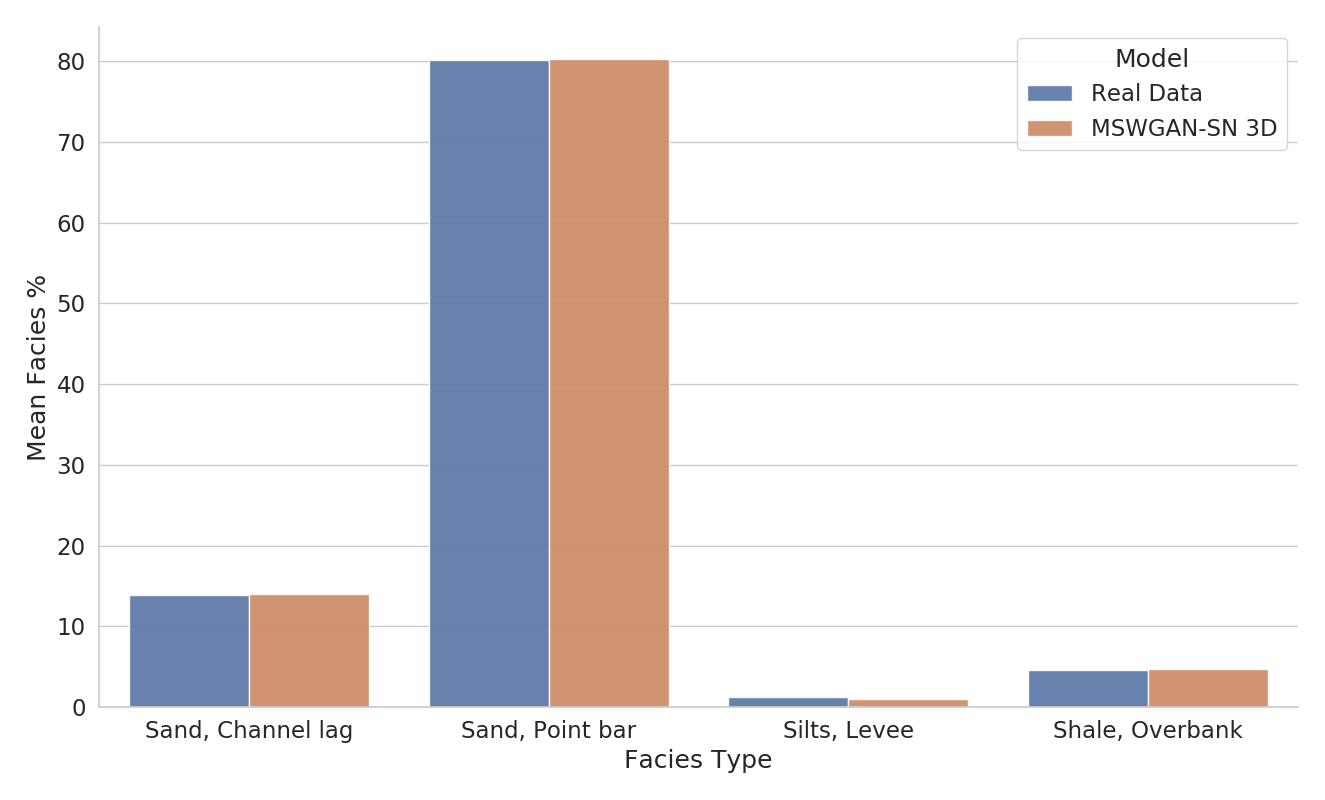}
    \captionof{figure}{Comparison of the proportion of different facies for our models.}
    \label{fig:3d_proportions}
\end{center}

\begin{center}
    \includegraphics[width=\linewidth]{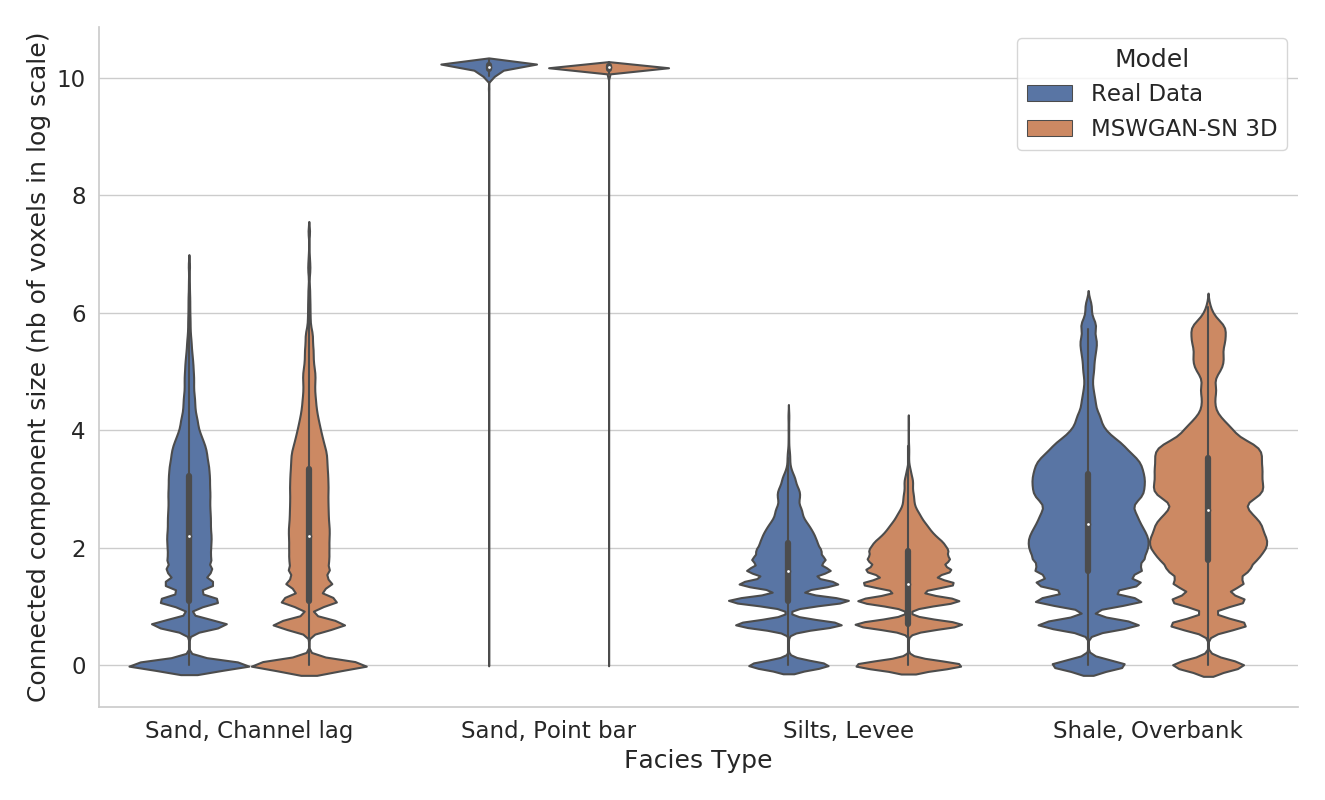}
    \captionof{figure}{Distribution of connected components size of each geological facies for the 3D model
against training data. To make visualisation more relevant, each connected component size is itself
weighted by it’s volume.}
    \label{fig:3d_connected_components}
\end{center}

\begin{table}[h]
\begin{center}
\begin{tabular}{| c | c  | c | c |}
 \hline
 Sand, Channel lag & Sand, Point bar & Silts, Levee & Shale, Overbank \\ 
\hline
 0.24 & 559.96 & 0.28 & 0.90 \\
\hline
\end{tabular}
\end{center}
\caption{In Figure \ref{fig:3d_connected_components}, we plot the distribution of connected components. To facilitate the comparison between each distribution, we compute a Wasserstein-1 distance between each of these distributions and the real distribution.}
\label{tab:connected_components3d}
\end{table}

\subsection{Parameterized Model}
\label{sec:parametrisation}

In order to capture the large range of geometries in channelized reservoir models, Flumy proposes two key parameters: the Net to Gross (sand proportion) and the sand bodies extension (lateral extension of point bars). Using these two parameters, the user can generate a large variety of different reservoirs (see Figure 7 in \cite{flumy}). 

The GAN framework can be upgraded such that it can be parameterized with some extra information $\phi$ \citep{mirza2014conditional}. This complementary information is inputted to both the generator and the critic. As such, the generator learns how to generate a realisation $x = G(z|\phi)$ and the critic learns to give a probability $D(x|\phi)$ that the realisation is real and corresponds to parameters $\phi$. Consequently, the payoff in this new framework is:
\begin{equation}
    v(\theta^G, \theta^D) = \mathbb{E}_{x\sim p_r} \log\ D(x|\phi) + \mathbb{E}_{x\sim p_{\theta}} \log(1-D(x|\phi))
\end{equation}

The problem with the original conditional GAN is that it is not translation invariant, because the architecture includes fully connected layers. Instead, we developed a stationary conditional GAN framework. Our approach does not change the output of linear layers but instead, learns different convolution kernels for each pair of input parameters.

This is done by adding a linear layer that will transform the parameters to an embedding space, which is then given to a transposed convolution layer. The outputs of these two layers are depth-wise convolution filters that can be applied to matrices. These layers and their parameterized filters are present in the generator and the critic. As such, both networks now have a second input in addition to their normal input. These two layers are trained alongside the generator and critic with the same gradient. In this approach, the dataset now consists of Flumy realisations from multiple distributions, corresponding to different parameter values. 
The generator is given random parameter values within the range of the real parameters when trained, allowing it to interpolate between parameter values not present in the training data.

\begin{center}
\includegraphics[width=0.75\linewidth]{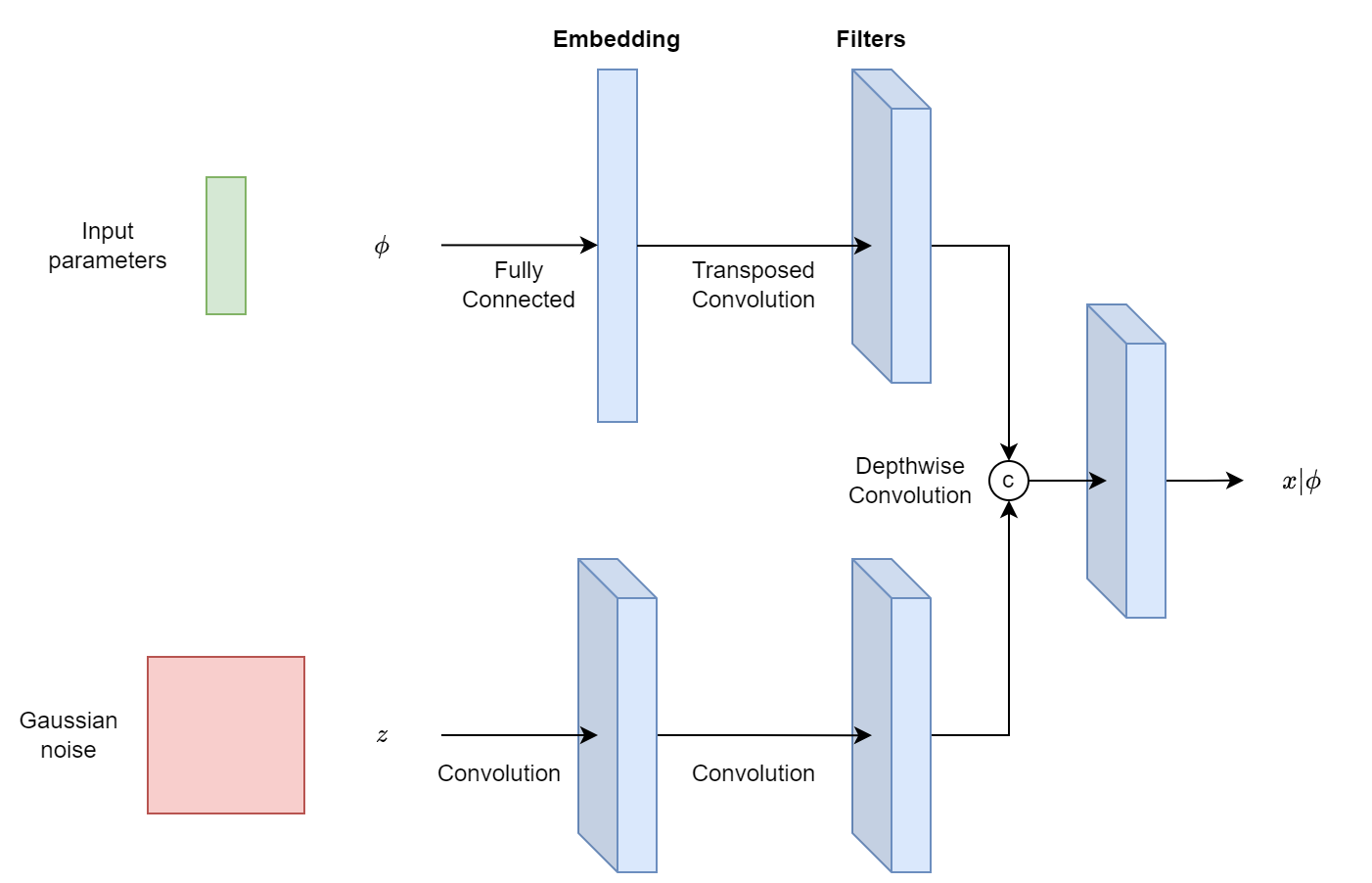}
\captionof{figure}{Diagram of the stationary conditional generator, where the input parameters $\phi$ are two scalar values in a vector (Net to Gross and Sand Body Extension), which are then projected into an embedding space before computing convolution kernels that are used in the network.}
\end{center}

\begin{center}
\includegraphics[width=\linewidth]{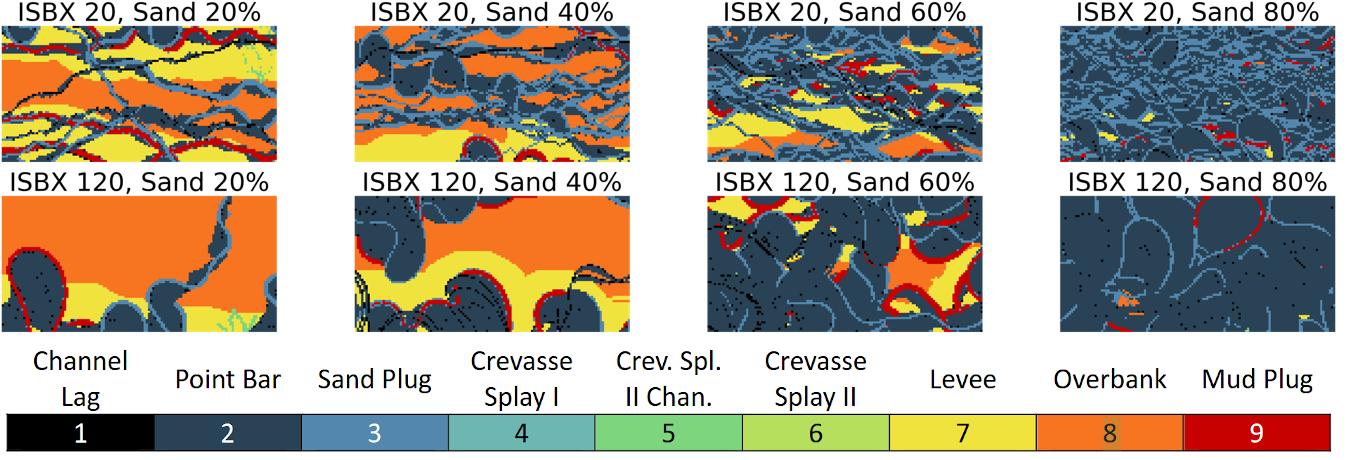}
\captionof{figure}{Realisations generated with Flumy by varying two parameters: the Net to Gross (NG, eg. the required Sand Proportion) and the Sand Body Extension (ISBX). Flow direction is from left to right. The goal of parametrization is to capture the relationship between the realizations and the input parameters.}
\label{fig:paramestrisation_dataset}
\end{center}

\begin{center}
\includegraphics[width=\linewidth]{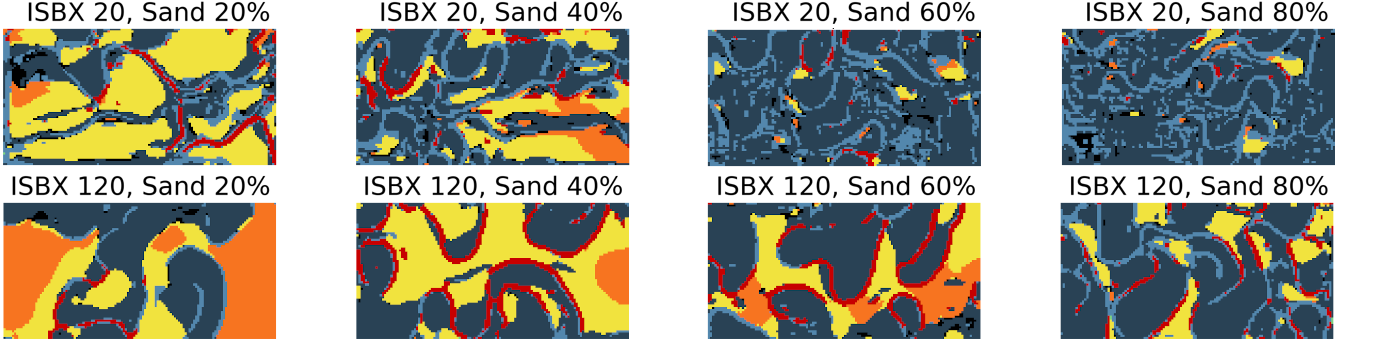}
\captionof{figure}{After training, the CGAN model can generate new realisations, taking into account the parameters it was inputted with.}
\label{fig:paramestrisation_generated}
\end{center}
Pictured in Figure \ref{fig:paramestrisation_dataset} are the simulations, generated with Flumy using a range of ISBX and NG values, used to train our CGAN experiments . Figure \ref{fig:paramestrisation_generated} displays the results of the trained CGAN model.
As can be seen from the above figures, the CGAN models can be modified to learn the relation between parameters and output realizations. The generator learns to account for these parameters, making it a useful feature for those wishing to use it as a proxy for Flumy.

\subsection{Conditional Generation}
\subsubsection{2D}

\begin{center}
\includegraphics[width=\linewidth]{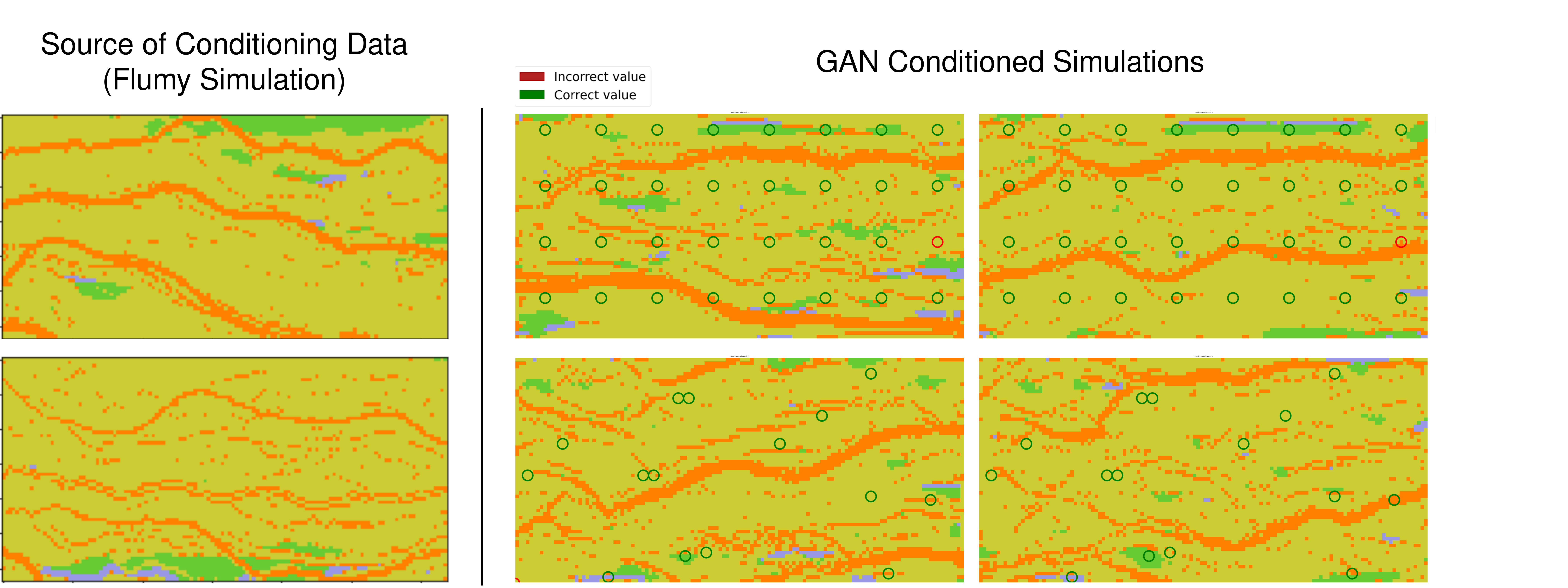}
\captionof{figure}{Example of conditional simulations from (circles, red for an error, green for a correct prediction). Top simulations were conditioned with 32 pixels sampled in a grid-like pattern from a Flumy simulation (leftmost). Bottom simulations were conditioned with 15 pixels sampled in a random pattern from another Flumy simulation (leftmost).}\label{conditioning_pixels}
\end{center}

Conditional generation can be tested by sampling Flumy simulations and randomly selecting vertical wells locations and masking the rest. In 2D, wells are pixels in the realisation where the category of facies is known. Similarly to the non-conditional methodology, we first test the method in two-dimension and only after in three-dimension. We evaluate how strictly the constraints are satisfied using the percentage of correctly generated values at the well. As shown in the Figure \ref{conditioning_percentage_2d}, the neural network adherence to the observation is marginally superior to that of the Gaussian mixture method.\\
However, assessing the extent to which the estimated posterior captures the variability of the real posterior is non-trivial. Instead, we judge how well it is captured by our network by comparing it to rejection sampling.
We generate thousands of simulations with the non-conditional generator, and reject any realisation that do not fulfill the given conditions. We then sample 2500 simulations using the two methods. We finally compare the probability at each pixel of every facies type between the two methods. This is shown in Figure \ref{conditioning_proba}. The results showcased highlight that the inference neural network does not seem to explore the whole domain of possible conditional realisations. In the top row, the probabilities of the first facies are only close to one in the immediate neighborhood of the conditional points, and there are a few pixels with a zero probability. This indicates that the sand channels have no particular trajectories in the rejection-sampling conditional distribution. However, the learned approximation of the conditional distribution exhibits over-confidence on a small set of sand channel trajectories, which are almost always sampled (in red in the sand facies probability maps). To address this issue, as mentioned in Section \ref{gaussian_mixture_part}, the neural network model is replaced by a Gaussian Mixture. As can be seen in Figure \ref{conditioning_proba}, this method  generates a probability map closer to that of the accept-reject method. Indeed, the sand channels do not seem to be constrained to a limited number of trajectories.

\begin{center}
\includegraphics[width=\linewidth]{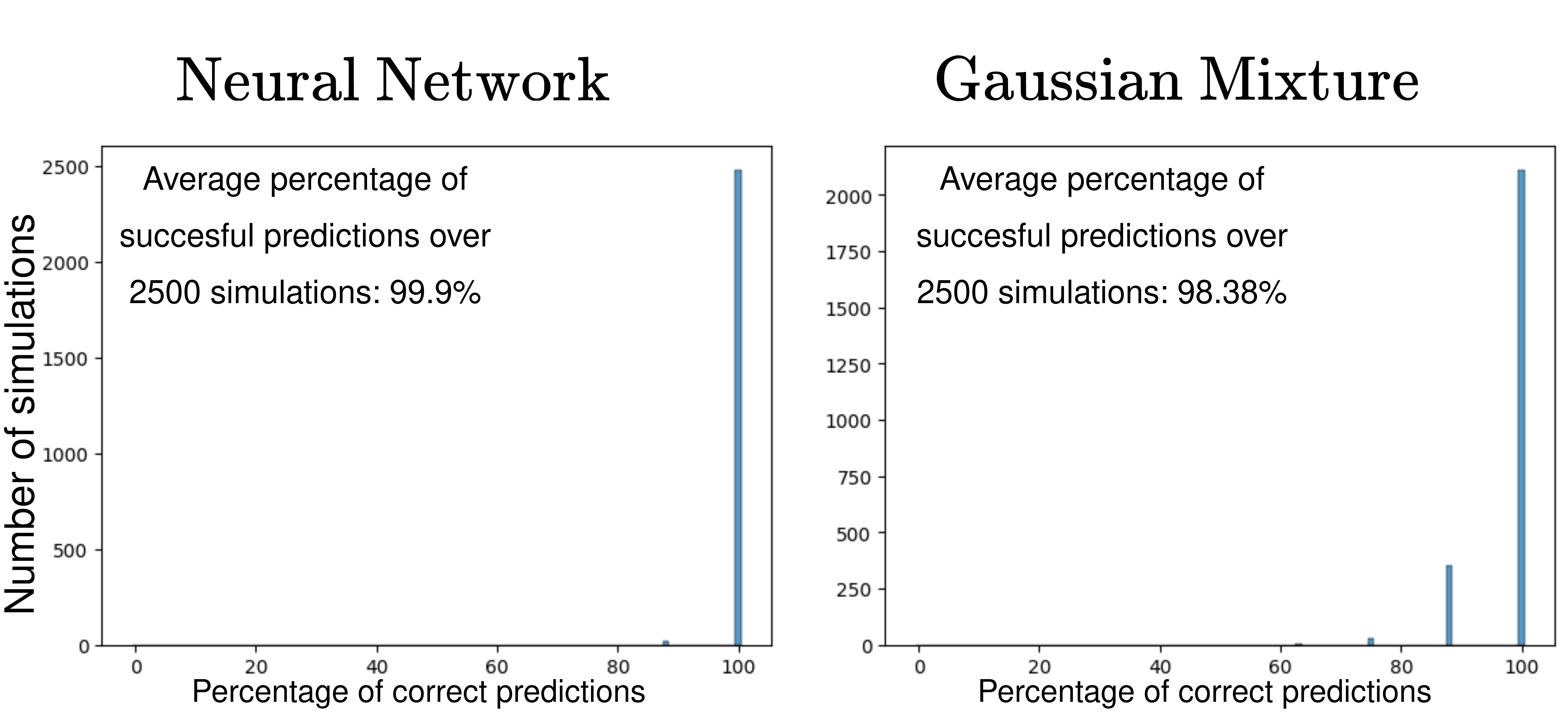}
\captionof{figure}{Comparison of the percentage of success using our two conditioning methods.}\label{conditioning_percentage_2d}
\end{center}

\begin{center}
\includegraphics[width=\linewidth]{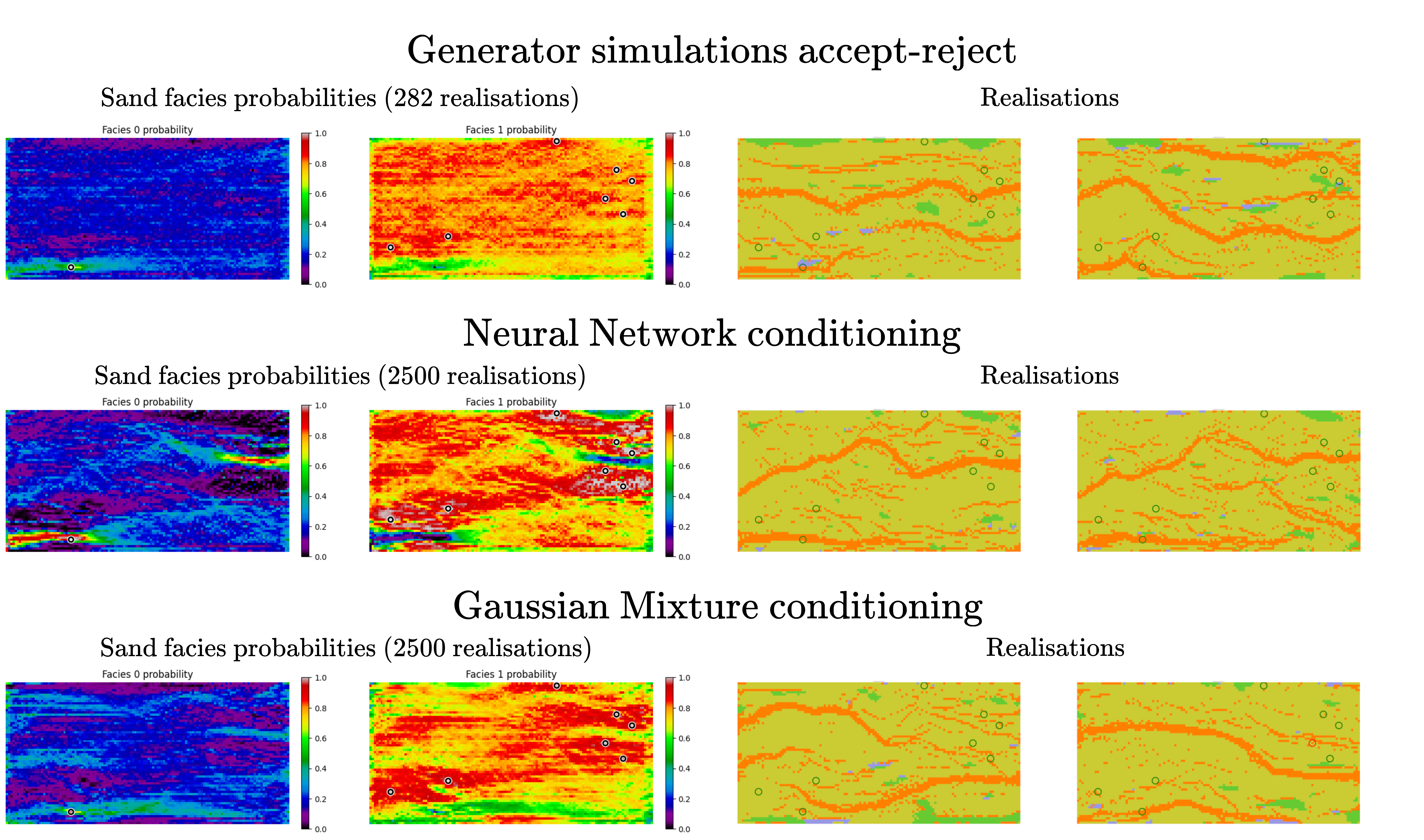}
\captionof{figure}{After the inference model is trained, we can generate any number of realizations. Here we generate 2500 realizations, we compute at each coordinate the probability of occurrence for each facies, and we visualize the result (left), as well as sampling two simulations for each methods (right). We compare accept-reject with the generator, neural network conditioning model and Gaussian mixture conditioning model probabilities.}\label{conditioning_proba}
\end{center}

\subsubsection{3D}
Both conditioning methods are also applicable and work in 3D, where observation wells correspond to a line of voxels from the top of the volume to the bottom. Some results are shown in Figure \ref{conditioning_results_3d}.
In 3D, the differences between the two methods are more blatant. First, the neural network method exhibits, on average, a better match to the conditioning data compared to the mixture method, as shown in Figure \ref{conditioning_percentage_3d}. This is a clear advantage of the large expressive power offered by deep learning algorithms. However, the difference in variability is also more pronounced in higher dimension, as seen in Figure \ref{conditioning_proba_3d}. This highlights the advantage of the mixture model to capture a broader range of variations.
We have found that 10 is an adequate number of components, enough to exhibit diversity while keeping computation cost as low as possible.
Indeed, the main disadvantage of the Gaussian mixture is the training time. Indeed, the covariance matrix of each component is of size $d\times d$, with $d$ the dimension of the latent space.  The algorithm also needs a huge batch size to ensure that every components of the mixture is sufficiently trained at each iteration (e.g., 350 in our experiments).

\begin{center}
\includegraphics[width=\linewidth]{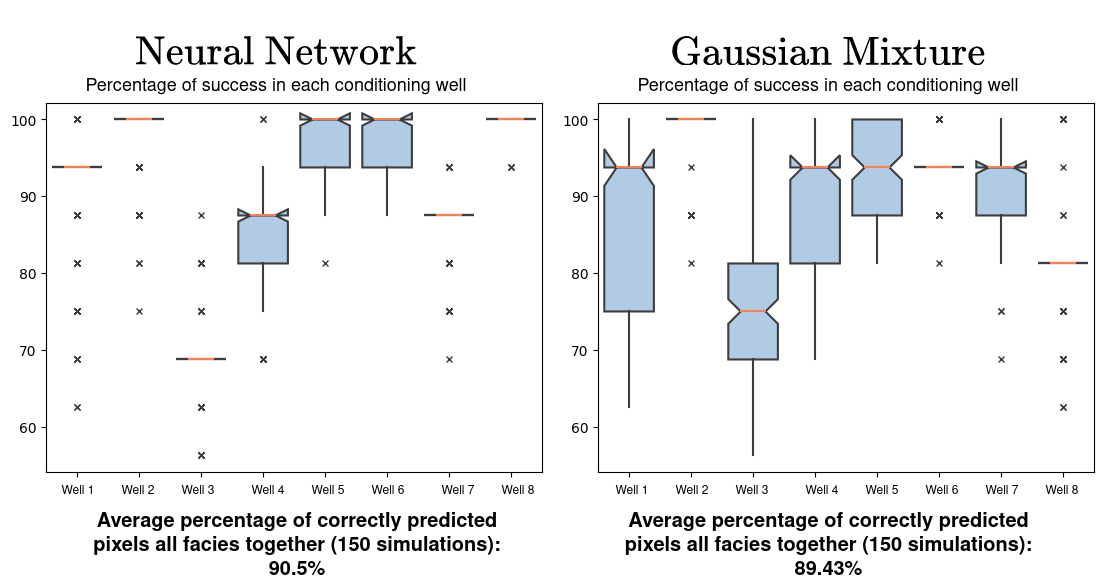}
\captionof{figure}{Comparison of the percentage of success, per well, using our two conditioning methods in 3D. The orange line is the median of the set and the crosses are outliers.}\label{conditioning_percentage_3d}
\end{center}

\begin{center}
\includegraphics[width=\linewidth]{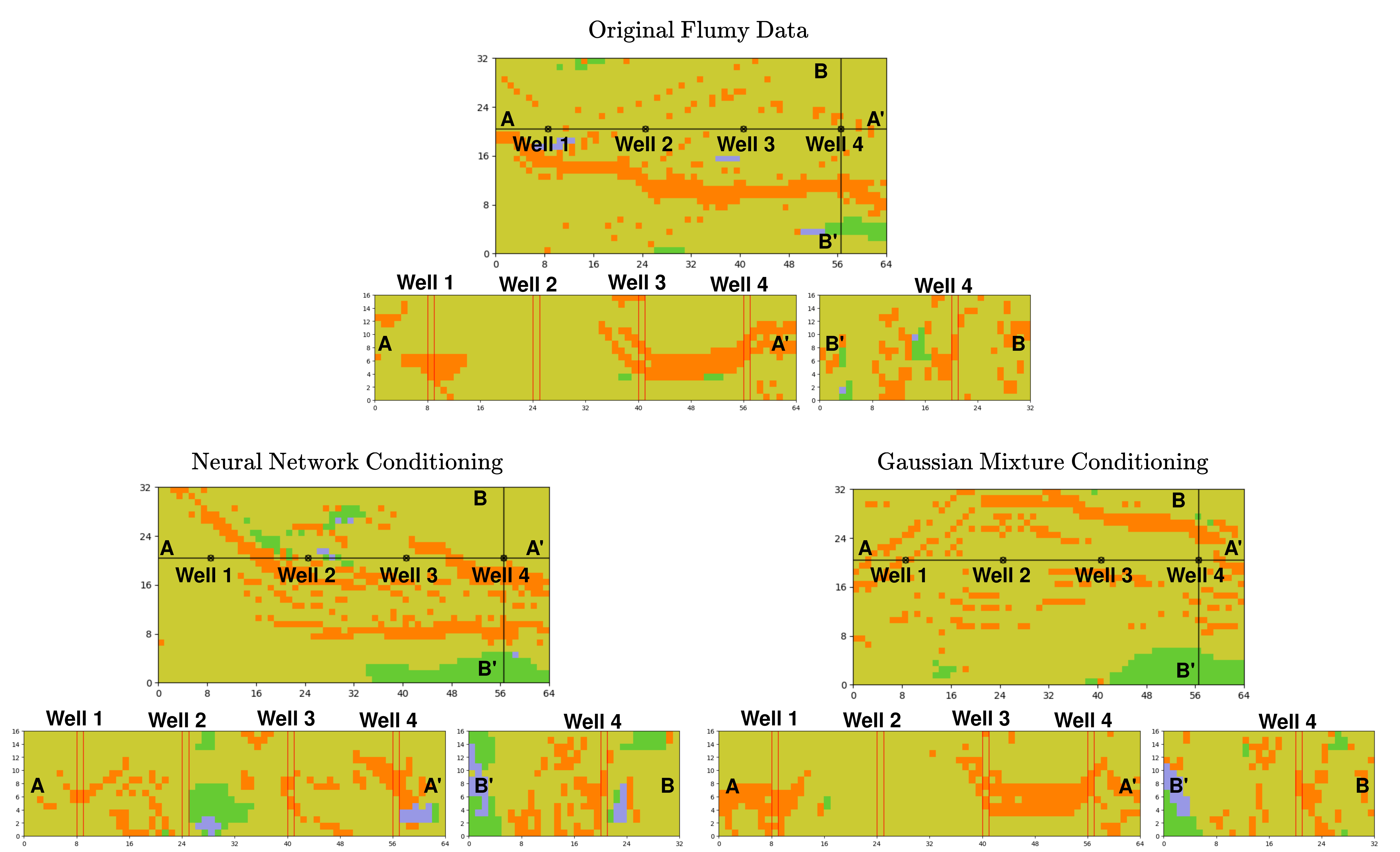}
\captionof{figure}{Conditional simulations with both methods trained in 3D (bottom) compared to the original Flumy volume (top) where the observations were taken. For each model three slices are shown: The upper one is an horizontal slice of the volume, on the bottom left and bottom right, the vertical sections $A$-$A'$ and $B'$-$B$, respectively. }\label{conditioning_results_3d}
\end{center}

\begin{center}
\includegraphics[width=\linewidth]{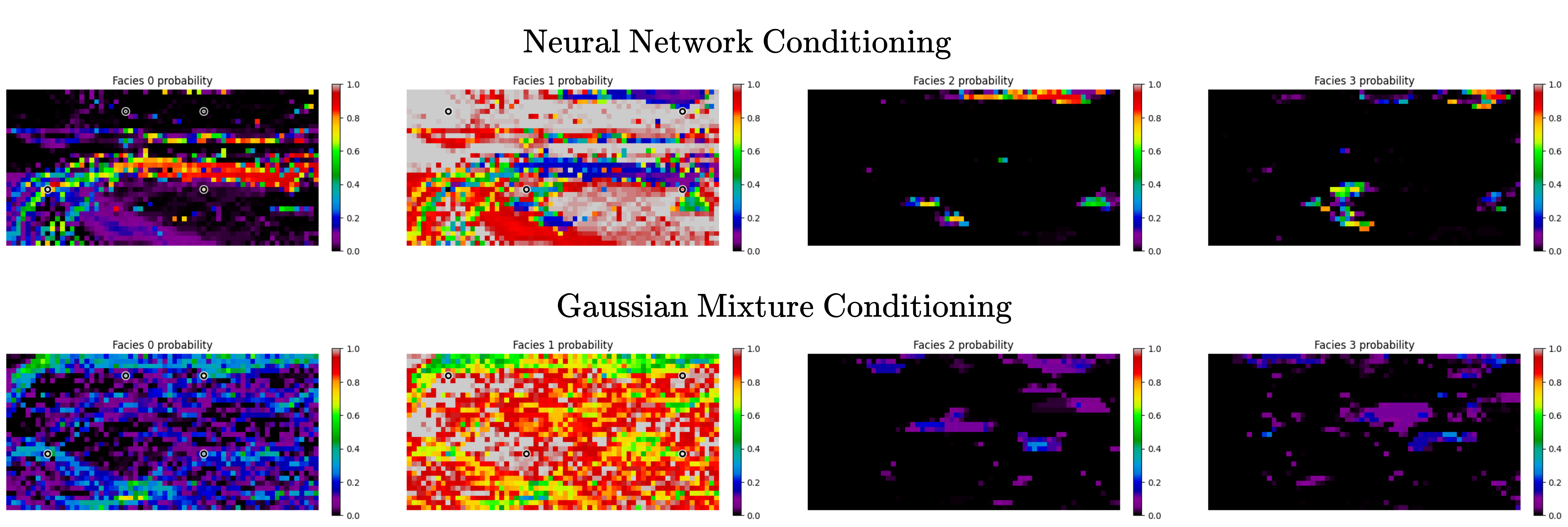}
\captionof{figure}{Identically to the 2D case, we can generate any number of simulations in 3D. Here we generate 150 realizations, we compute at each coordinate the probability of each facies occurring, and we visualize the results of a selected horizontal slice of the 3D volume.}\label{conditioning_proba_3d}
\end{center}

\section{Conclusion}
In this paper, we presented a deep generative model, as well as a method to condition it to observations. Multiple elements of the recent literature are incorporated in our non-conditional architecture, in order to make it robust enough to capture the characteristics of datasets generated with the Flumy Model. We discussed both theoretical and empirical results that support the applicability of our proposed approach for simulating conditional meandering channelized reservoirs. Our experimental results indicate that this approach is successful, in 2D and 3D, as an alternative to Flumy to generate conditional geological facies configurations. However, it should be noted that we limited most of our experiments to a specific scenario with a high Net to Gross having 4 facies, and further experimentation needs to be done to establish the applicability of the conditioning method to more complex Flumy scenarios.

The results of the present study suggest that the method is applicable to subsurface facies characterization. The non-conditional design choices in architecture seem to mitigate the usual GAN flaws described in the literature. This makes the method applicable to different datasets.
The conditional simulations method may lead to a poor solution with inconsistencies with the constraints, especially in 3D. Because this is a statistical method, we cannot always guarantee that the constraints will be fulfilled in all simulations. However, we also believe that this happens because, by construction, the conditional method relies on the assumption that our prior generator model is able to generate any simulation. Therefore, the method's performance could be improved by using a prior model with more expressive power. 

This variational conditioning scheme is adaptable such that future research could focus on conditioning to seismic imagery or history matching. Seismic imagery seems the most natural continuation when considering the similarity between seismic proportion maps and low resolution realisations. However, to improve the method, the inference model should be capable of taking advantage of the expressive power and speed of neural networks while being able to have the variability of mixtures.

Applications for the proposed method include resource prospecting and meteorological simulations. Some applications may require to test the constraints against a range of simulations generated with various parameters. This motivated us to successfully investigate the use of a conditional GAN.

All of this suggests promising future work towards even more realistic and flexible models. We will take advantage of the robustness of our models to generate spatio-temporal data using Transformer layers \citep{attentionisallyouneed}.

\section{Acknowledgments}

This method has been developed within the scope of the Flumy Research Program (JIP).
The authors are grateful to ENI partners for support and fruitful discussions.



\newpage

\textbf{Code availability section}

A stable deep adversarial learning approach for geological facies generation

Contact: ferdinand.bhavsar@minesparis.psl.eu

Hardware requirements: A GPU with CUDA library installed is highly recommended. 16GB of System RAM, as well as GPU RAM, is recommended for 3D codes.

Program language: Python
 
Software required: see requirements.txt

Program size: 57.1MB (+1.2GB of data on the git repository)

The source codes are available for downloading at the link:
https://github.com/Pumafi/principled\_dlgeo

\bibliographystyle{cas-model2-names}
\bibliography{biblio.bib} 

\appendix

\section{Final Model Architecture}
\label{appendix:MSWGAN_arcitecture}
\begin{center}
\includegraphics[width=1\linewidth]{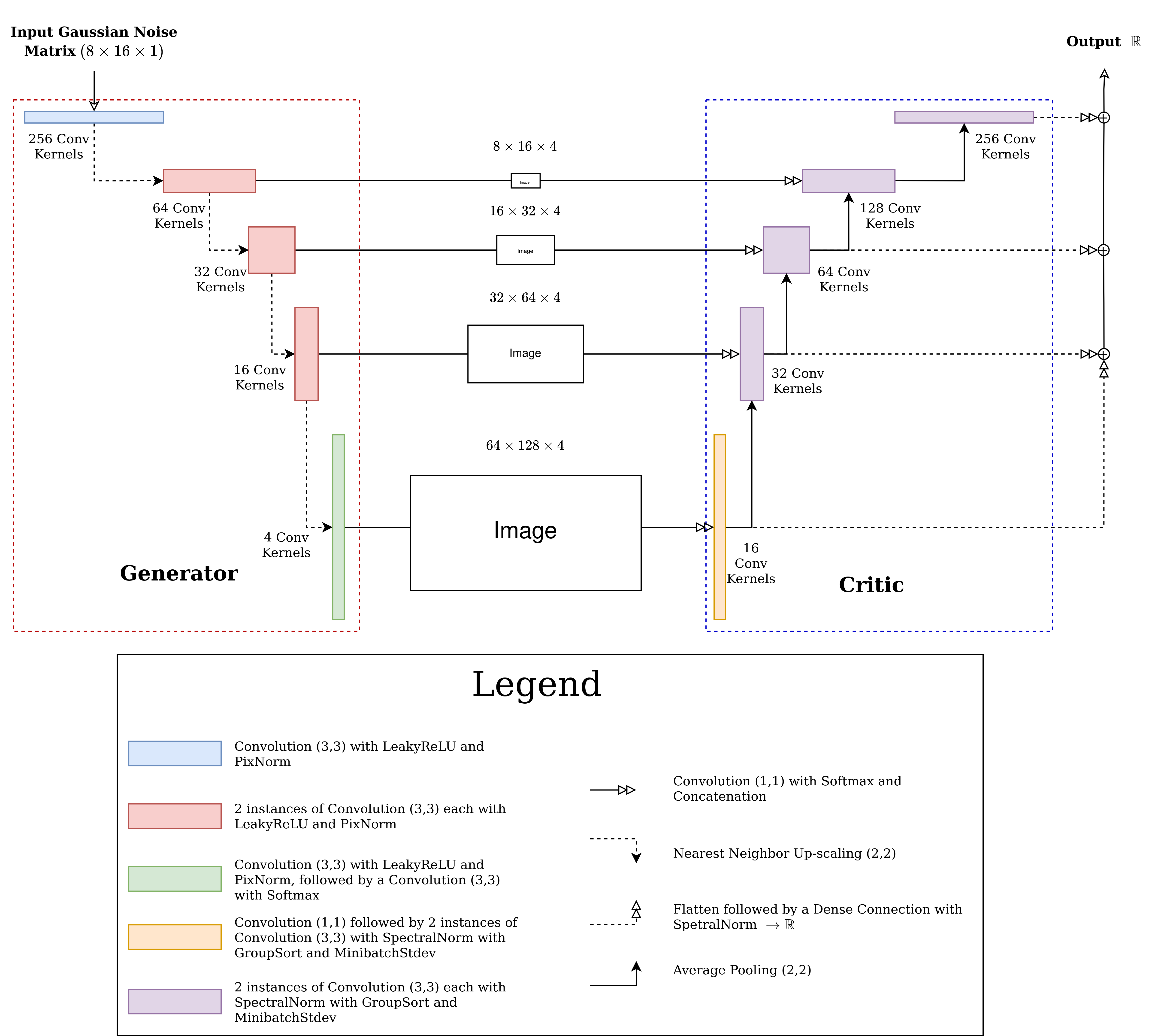}
\captionof{figure}{Diagram of the final model adapted from pre-existing methods and tests}
\end{center}

\section{Detailed Architecture Design and Mathematics}
\subsection{Wasserstein-GAN and Kantorovich-Rubinstein duality}
\label{appendix_kr_dual}
Instead of making the overlap more likely, another popular method avoid the problem completely by resorting to another loss function. The Wasserstein-1 distance, a distance measure between two probability distributions, has been proposed \citep{arjovsky2017wasserstein}. Unlike the Jensen-Shannon divergence, the derivative of the Wasserstein distance is never constant, such that there  is always a usable gradient, making it an ideal candidate to improve the robustness of the networks. 
It was first defined for optimal transport problems, under the name Earth-Mover distance, to define the minimum effort to move an earth-pile to a hole. There are infinite transport plans to move the earth-pile, but not all are optimal. By changing the earth-pile and hole to probability distributions and the transport plans to joint distributions, we derive a loss function to train a GAN \citep{arjovsky2017wasserstein}. Our distributions are the ones defined by the generator model with parameters $\theta$ and the training data distribution. They are respectively named $p_\theta$ and $p_r$ and are defined on the same probability space $M$.

A transport plan is noted $\gamma(x, y) $. It is a joint distribution with $p_{\theta}$ and $p_{r}$ as marginals. The Wasserstein distance is defined as the expectation of the euclidean distance between pairs of random variables following this joint distribution:\\

\begin{equation}
W(p_{r}, p_{\theta}) = \underset{\gamma \in \Pi}{\inf}\ \underset{(x,y) \sim \gamma}{\mathbf{E}}  ||x-y||_1
\end{equation}

We cannot study all joint probability distributions to find the optimum, as the optimisation problem is too complex. Instead, we use the simpler form of the Kantorovich-Rubinstein dual \citep{villani_topics_2003, villani_optimal_2008}.
\begin{theorem}
\label{appendix_kr_dual_th}
Let $p_r$ and $p_\theta$ be two probability distributions defined on a probability space $M$. Let $\Pi$ be the set of all joint distributions such that $p_r$ and $p_\theta$ are the marginal distributions. Let $\gamma$ be an element of $\Pi$ and let $(x, y)$ be realizations of $\gamma$. The Wasserstein divergence between $p_r$ and $p_\theta$ can be expressed in the following dual form:
\begin{equation}
    W_1(p_{r}, p_{\theta})  = \underset{\gamma \in \Pi}{\inf}\ \underset{(x,y) \sim \gamma}{\mathbf{E}}  ||x-y||_1 = \underset{\substack{f\\Lip(f) \leq 1 }}{\sup} \underset{x \thicksim P_r}{\mathbf{E}}f(x)  - \underset{y \thicksim P_\theta}{\mathbf{E}}f(y)
\end{equation}
\end{theorem}

where the supremum is taken over all functions $f$ that are 1-Lipschitz, which is the property that for any $x$ and $y$, $f(x) + f(y) \leq ||x-y||_1$. 

We take the Lagrangian associated with the constrained equation, with Lagrange multipliers $f, g: M \xrightarrow[]{} \mathbb{R}$ (\textbf{continuous in $M$}) \citep{villani_topics_2003, villani_optimal_2008}, using the constraints on the transport plan:
\begin{equation}
\begin{aligned}
\mathcal{L}(\gamma, f, g) = {} & \underset{(x,y) \sim \gamma}{\mathbf{E}}  ||x-y||_1 + \int_M\Big(P_r(x) - \int_M\gamma(x,y)dy\Big)f(x)dx\\
& + \int_M\Big(P_\theta - \int_M\gamma(x,y)dx\Big)g(y)dy
\end{aligned}
\end{equation}

When $P_r(x) = \int_M\gamma(x,y)dy$ and $P_\theta(y) = \int_M\gamma(x,y)dx$ then we have $\mathcal{L}(\gamma, f, g) = {\mathbf{E}}  ||x-y||_1$.\\
However when $P_r(x) \neq \int_M\gamma(x,y)dy$ or $P_\theta(y) \neq \int_M\gamma(x,y)dx$ then ${\sup}\ \mathcal{L}(\gamma, f, g) = +\infty$.\\
Therefore, the optimisation problem is equivalent to a min-max problem:
\begin{equation}
W_1(P_{r}, P_{\theta}) = \underset{\gamma \in \Pi}{\inf}\ \underset{(x,y) \sim \gamma}{\mathbf{E}}  ||x-y||_1 = \underset{\gamma}{\inf}\ \underset{f, g}{\sup}\  \mathcal{L}(\gamma, f, g)
\end{equation}

We can write the Lagrangian in the following form:
\begin{equation}
\begin{aligned}
\mathcal{L}(\gamma, f, g) = {} & \underset{(x,y) \sim \gamma}{\mathbf{E}}  ||x-y||_1 + \int_M\Big(P_r(x) - \int_M\gamma(x,y)dy\Big)f(x)dx + \int_M\Big(P_\theta - \int_M\gamma(x,y)dx\Big)g(y)dy\\
& = \int_{M\times M} ||x-y||_1\gamma(x,y)dxdy + \int_M P_r(x)f(x)dx - \int_{M\times M}f(x)\gamma(x,y)dydx\\
& + \int_MP_\theta(y)g(y)-\int_{Mes M}g(y)\gamma(x,y)dxdy\\
& = \underset{x \thicksim P_r}{\mathbf{E}}f(x)  + \underset{y \thicksim P_\theta}{\mathbf{E}}g(y) + \int_{M\times M} \Big(||x-y||_1 - f(x) - g(y)\Big)\gamma(x,y) dxdy
\end{aligned}
\end{equation}

The Strong Duality theorem states that if an optimisation problem is linear, and the primal and dual form both are realisable, i.e. their constraints are not incompatible, then the optimum of both forms are equal. We can use this theorem because we are in a linear optimisation problem and constraints are compatible. Therefore, we rewrite the equation and switch $\sup$ and $\inf$.
\begin{equation}
    \begin{aligned}
  W_1(P_{r}, P_{\theta}) = \underset{\gamma}{\inf}\ \underset{f, g}{\sup}\  \mathcal{L}(\gamma, f, g)\ = {} & \underset{f, g}{\sup}\ \underset{\gamma}{\inf}\  \mathcal{L}(\gamma, f, g)
  \end{aligned}
\end{equation}

If  $||x-y||_1 < f^*(x)+g^*(y)$, the $\inf$ of the Lagrangian goes to $-\infty$, while if $||x-y||_1 \geq f^*(x)+g^*(y)$, then $\underset{\gamma}{\inf}\mathcal{L}(\gamma, f, g) = 0$. Therefore we can write:
\begin{equation}
\underset{\gamma}{\inf}\int_{M\times M} \Big(||x-y||_1 - f(x) - g(y)\Big)\gamma(x,y) dxdy =  \begin{cases}
      0 & \text{si $||x-y||_1 \geq f(x)+g(y)$} \\
      -\infty & \text{else}
    \end{cases}
\end{equation}

Working on this Lagragian allows us to ignore the constraints on $\gamma$:
\begin{equation}
     \int_{M\times M} \Big(||x-y||_1 - f(x) - g(y)\Big)\gamma(x,y) dxdy
     = \int_{M\times M} F(x,y)\gamma(x,y) dxdy
\end{equation}

With $F(x,y) = (||x-y||_1 - f(x) - g(y))$, $F(x,y): M\times M \xrightarrow{} \mathbb{R}$, $F$ continuous on its domain as the sum of continuous functions.\\
If there is a $(x, y)$ such that $F(x, y) < 0$ then, because $F$ is continuous, there exist an interval $I \in M \times M$ where the function takes negative values. This interval is the neighbourhood of $(x,y)$.\\
Let's take $F$ such that on interval $I$ the function takes values in $[-\epsilon, -\epsilon/2]$, $\epsilon > 0$, and takes the value $0$ on the rest of its domain. We can then write:\\
\begin{equation}
\begin{aligned}
    \int_{M\times M} F(x,y)\gamma(x,y) dxdy = \int_{(M\times M) / I} F(x,y)\gamma(x,y) dxdy + \int_{I} F(x,y)\gamma(x,y) dxdy\\
    & = \int_{I} F(x,y)\gamma(x,y) dxdy \\
    & \leq -\epsilon/2\int_{I} \gamma(x,y) dxdy
\end{aligned}
\end{equation}

By giving $\gamma(x,y)$ an arbitrarily large value on $I$, we make this integral tend towards $-\infty$. Therefore, when $||x-y||_1 < f^*(x)+g^*(y)$, the $\inf$ of the Lagrangian tends towards $-\infty$. We then rewrite this part of the equation as a constraint on $f$ and $g$:
\begin{equation}
    \begin{aligned}
  W_1(P_{r}, P_{\theta}) = \underset{\substack{f, g\\||x-y||_1 \geq f(x)+g(y) }}{\sup} \underset{x \thicksim P_r}{\mathbf{E}}f(x)  + \underset{y \thicksim P_\theta}{\mathbf{E}}g(y)
  \end{aligned}
\end{equation}

The problem is now to find the optimal $f$ and the optimal $g$, i.e. maximise this equation with constraint $||x-y||_1 \geq f(x)+g(y)$.

We can show that $g(x) = -f(x)$:
\begin{equation}
    ||x-y||_1 \geq f(x)+g(y)
\end{equation}
Since we take the $sup$ over all function $f$ and $g$ for which the condition is true, we can determine that: 
\begin{equation}
    ||x-y||_1 = f(x)+g(y)
\end{equation}
This equation is true $\forall x, \forall y$, therefore when $x = y$:
\begin{equation}
    \begin{aligned}
    f(x)+g(y) = ||x-y||_1 = 0
  \end{aligned}
\end{equation}
Therefore $g(x) = -f(x)$, which gives the following formula:
\begin{equation}
    \begin{aligned}
  W_1(P_{r}, P_{\theta}) = \underset{\substack{f\\||x-y||_1 \geq f(x)-f(y) }}{\sup} \underset{x \thicksim P_r}{\mathbf{E}}f(x)  - \underset{y \thicksim P_\theta}{\mathbf{E}}f(y)\\
  \Leftrightarrow   W_1(P_{r}, P_{\theta}) = \underset{\substack{f\\Lip(f) \leq 1 }}{\sup} \underset{x \thicksim P_r}{\mathbf{E}}f(x)  - \underset{y \thicksim P_\theta}{\mathbf{E}}f(y)
  \end{aligned}
\end{equation}

 We use this new formulation of the distance to replace the loss function in the previously defined GAN architecture. The 1-Lipschitz function $f$ of the formula is the function defined by our discriminator. Therefore, two major changes are imposed on the architecture:
\begin{enumerate}
    \item Since $f$ can take any value on $\mathbb{R}$, the last sigmoid activation function present in the original GAN architecture \citep{goodfellow2014generative} is removed.
    \item The discriminator is constrained to maintain its Lipschitz bound inferior or equal to $1$.
\end{enumerate}

In the original Wasserstein GAN article \citep{arjovsky2017wasserstein}, the 1-Lipschitz constraint is enforced through gradient clipping. 
Gradient clipping simply controls the Lipschitz norm of the model by re-scaling the norm of the gradient to be at most $1$. Any gradient norm exceeding this value is re-scaled accordingly.
Later studies pointed out that gradient clipping leads to the undesirable behaviour of capacity underuse, leading to the critic ending up learning extremely simple functions \citep{gulrajani2017improved}. These over-simplified functions are a problem, as they might fail to represent the Wasserstein loss correctly. As such we leverage properties on the norm of the gradient (see Annex \ref{appendix_proba_lip_dual} and \ref{appendix_opi_cri_grad}) and use Spectral Normalization (see Section \ref{descript_gan_archi}).

\subsection{Lipschitz norm of Kantorovich-Rubistein dual}
\label{appendix_proba_lip_dual}
\begin{theorem}
\label{lipchitz_norm_th}
Given $\gamma^*$ the optimal transport plan minimising: $\underset{\gamma \in \Pi}{\inf}\  \underset{(x,y) \sim \gamma}{\mathbb{E}} ||x-y||_1$. If  $f^*$, the solution of the Kantorovich-Rubinstein dual (also called optimal critic) is differentiable, then we have the following equality:\\
\begin{equation}
    P[f^*(x)-f^*(y) = ||x-y||_1] = 1
\end{equation}
\end{theorem}

Thanks to Theorem \ref{appendix_kr_dual_th} we can write:
\begin{equation}
\underset{||f||_L \leq 1}{\sup}\ \underset{x \sim P_r}{\mathbb{E}} f(x) - \underset{y \sim P_\theta}{\mathbb{E}} f(y) = \underset{\gamma \in \Pi}{\inf}\  \underset{(x,y) \sim \gamma}{\mathbf{E}} ||x-y||_1
\end{equation}

Because of the Strong Duality theorem we know both optimal values are equal, and we can write:
\begin{equation}
\begin{aligned}
{} & \underset{(x,y) \sim \gamma^*}{\mathbf{E}} \Big[ f^*(x) -  f^*(y) \Big] = \underset{(x,y) \sim \gamma^*}{\mathbf{E}} ||x-y||_1\\
\Leftrightarrow & \underset{(x,y) \sim \gamma^*}{\mathbf{E}} \Big[ f^*(x) -  f^*(y) \Big] - \underset{(x,y) \sim \gamma^*}{\mathbf{E}} ||x-y||_1 = 0 \\
\Leftrightarrow & \underset{(x,y) \sim \gamma^*}{\mathbf{E}} \Big[ f^*(x) -  f^*(y) - ||x-y||_1 \Big] = 0
\end{aligned}
\end{equation}

The expectation is zero, therefore the probability distribution is centered. But since the function is 1-Lipschitz continuous, we can write $f^*(y)-f^*(x) \leq ||y - x||_1$. Both properties are only possible if all the mass of the probability distribution is at $0$.

\subsection{Optimal critic gradient}
\label{appendix_opi_cri_grad}
\begin{theorem}
If $f^*: \mathbb{R}^n \xrightarrow{} \mathbb{R}$, $n \in \mathbb{N}$, is the 1-Lipschitz optimal critic that maximises the Kantorovich-Rubinstein dual, then $||\nabla f^*(x)||_1 = 1\ \text{almost everywhere}$.
\end{theorem}

Using the Frechet derivative generalisation to vector-valued functions we write the derivative of $f$:
\begin{equation}
\begin{aligned}
{} & \underset{||x - y||_1 \sim 0}{\lim} \frac{f^*(x) - f^*(y) - <\nabla f^*(x), y-x>}{||x - y||_1} = 0\\
\Leftrightarrow & \underset{||x - y||_1 \sim 0}{\lim} \frac{f^*(x) - f^*(y)}{||x - y||_1} = \underset{||x - y||_1 \sim 0}{\lim} \frac{<\nabla f^*(x), y-x>}{||x - y||_1} \\
\end{aligned}
\end{equation}

Using the Cauchy–Schwarz inequality we deduce a lower bound of this derivative:
\begin{equation}
\begin{aligned}
{} & \underset{||x - y||_1 \sim 0}{\lim} \frac{f^*(x) - f^*(y)}{||x - y||_1} = \underset{||x - y||_1 \sim 0}{\lim} \frac{<\nabla f^*(x), y-x>}{||x - y||_1} \leq \frac{||\nabla f^*(x)||_1.||y - x||_1}{||y - x||_1} = ||\nabla f^*(x)||_1\\
\end{aligned}
\end{equation}

Thanks to Theorem \ref{lipchitz_norm_th} we know that $P\Big[\frac{f^*(x)-f^*(y)}{||x - y||} = 1\Big] = 1$. Consequently $P\big[||\nabla f^*(x)||_1 \geq 1 \big] = 1$, i.e. $1$ is a lower bound of the gradient almost everywhere. Since the function is 1-Lipschitz, the upper bound of the gradient is also $1$. The upper and lower bound of the gradient are equal almost everywhere, therefore we proved that $||\nabla f^*(x)||_1 = 1\ \text{almost everywhere}$.

\cite{arjovsky2017principledtraining} uses gradient penalty to enforce this theorem locally on the critic. Enforcing it on the entirety of its domain of definition require to make every component of the critic 1-Lipschitz continuous.
Linear and Convolutional layers can be constrained to estimating 1-Lipschitz function by using Spectral Normalization \citep{spectral_normalisation}.

\subsection{Kozachenko-Leonenko Entropy Estimator}\label{appendix_kozachenko_leonenko}

Given a random variable $\Tilde{z}$ with unknown probability distribution $q_\theta$ defined by our inference network $I$, which we can sample. If we have $M$ realizations $\Tilde{z}_{i_{0 \leq i \leq M}}$, then for any $\Tilde{z}$ we can compute the $k$ nearest neighbours in the set of realization $\Tilde{z}_i$.
We note the distance from $\Tilde{z}$ to the $k^{th}$ neighbour $d_{knn}(\Tilde{z})$. $B(\Tilde{z}, d_{knn}(\Tilde{z}))$ is smallest ball with $\Tilde{z}$ as center containing its $k$ nearest neighbours. Its radius is $d_{knn}(\Tilde{z})$.

If we take the assumption that $q$ is continuous and $q(\Tilde{z})_{\forall \Tilde{z}}\approx q(y)_{\forall y \in B(\Tilde{z}, d_{knn}(\Tilde{z}))}$, then we can estimate the empirical Shannon entropy of $q_\theta$ using our $M$ realizations $\Tilde{z}_{i_{0 \leq i \leq M}}$, thanks to the Kozachenko-leonenko estimator\citep{kozachenkoleonenko}:
\begin{equation}
\label{eq_est_entropy}
\begin{aligned}
    \mathbf{\hat{H}}(q) = & \approx \log(M) - \log(k) + \log(V_{B(_, 1)}) + \frac{d}{M} \sum_{i=0}^{M} \log(d_{knn}(\Tilde{z}_i))
\end{aligned}
\end{equation}

Intuitively, what this formula means is that the closer is the $k^{th}$ nearest neighbour from $\Tilde{z}$, the lower is the entropy of $q_\theta(\Tilde{z})$.\\
Only $\frac{d}{M} \sum_{i=0}^{M} \log(d_{knn}(\Tilde{z}_i))$ is optimised in this equation. Therefore, we can replace the entropy in our loss function:
\begin{equation}
\label{eq:dkl_kozachenko}
\begin{aligned}
{} &\operatorname*{argmin}_{\theta} D_{KL}\Big(q_\theta(z|x^\star)\ ||\ p(z|x^\star)\Big)\\
& \approx \operatorname*{argmin}_{\theta}\ -\frac{d}{M} \sum_{i=0}^{M} \log(d_{knn}(z_i)) + \mathbb{E}_{z\sim q_\theta} \Big[-\sum_{i=1}^{N} \log p(x_i^{\star}|z) + \frac{1}{2}||z||^2\Big]
\end{aligned}
\end{equation}

\section{Data generation process}\label{data_gen_process}
The non-conditional Flumy reference simulation aims at producing meandering channelized reservoirs made by a small aggrading turbidite inside a canyon having a length of 1280m, a width of 640m and a height of 640m. The meandering turbidite channel has a maximum thickness of 10m, a constant width of 350m and a large wavelength of 4000m. The erodibility coefficient is reduced to 1e-08 in order to slow down lateral migration (low sinuosity). Aggradation events occur in average every 140 iterations with an intensity of 1m thickness, a small horizontal extension of 928m and a small levee width of 140m. Local avulsions are disabled and regional avulsions period is set to 110 iterations. This parameter set results in a simulation having a very high sand proportion with elongated channel lag footprints along the flow direction and some scattered shale lenses that can constitute flow barriers. 

\end{document}